\documentclass[iop,twocolappendix,numberedappendix,apj]{emulateapj}
\usepackage{epsfig, natbib, graphicx, color, amsmath, cancel,icomma,bm,eso-pic}

\AddToShipoutPictureBG*{%
  \AtPageUpperLeft{%
    \hspace{0.75\paperwidth}%
    \raisebox{-8.0\baselineskip}{%
      \makebox[0pt][l]{\textnormal{DES 2015-0146}}
}}}%

\AddToShipoutPictureBG*{%
  \AtPageUpperLeft{%
    \hspace{0.75\paperwidth}%
    \raisebox{-9.0\baselineskip}{%
      \makebox[0pt][l]{\textnormal{SLAC PUB-16454}}
}}}%

\AddToShipoutPictureBG*{%
  \AtPageUpperLeft{%
    \hspace{0.75\paperwidth}%
    \raisebox{-10.0\baselineskip}{%
      \makebox[0pt][l]{\textnormal{Fermilab PUB-16-012-E-PPD}}
}}}%

\AddToShipoutPictureBG*{%
  \AtPageUpperLeft{%
    \hspace{0.75\paperwidth}%
    \raisebox{-11.0\baselineskip}{%
      \makebox[0pt][l]{\textnormal{DOI 10.3847/0067-0049/224/1/1}}
}}}%

\newcommand{\redmapper}{redMaPPer}
\newcommand{\Redmapper}{RedMaPPer}
\newcommand{\redmagic}{redMaGiC}

\newcommand{\hMpc}{h^{-1}\,\mathrm{Mpc}}
\newcommand{\msun}{M_\odot}

\newcommand{\photoz}{photo-$z$}
\newcommand{\photozs}{photo-$z$s}
\newcommand{\pmem}{p_{\mathrm{mem}}}
\newcommand{\Pcen}{P_{\mathrm{cen}}}
\newcommand{\ngmix}{{\tt ngmix}}
\newcommand{\modest}{{\tt modest}}
\newcommand{\mangle}{{\tt MANGLE}}
\newcommand{\healpix}{{\tt HEALPIX}}
\newcommand{\nsidef}{{\tt NSIDE=4096}}
\newcommand{\magauto}{{\tt MAG\_AUTO}}
\newcommand{\pfree}{p_{\mathrm{free}}}
\newcommand{\Lthresh}{L_{\mathrm{thresh}}}
\newcommand{\nsamp}{n_{\mathrm{samp}}}
\newcommand{\nkeep}{n_{\mathrm{keep}}}

\newcommand{\be}{\begin{equation}}
\newcommand{\ee}{\end{equation}}
\newcommand{\bea}{\begin{eqnarray}}
\newcommand{\eea}{\end{eqnarray}}

\newcommand{\mi}{m_i}
\newcommand{\mz}{m_z}
\newcommand{\mzlim}{m_{z,\mathrm{lim}}}
\newcommand{\zlambda}{z_{\lambda}}
\newcommand{\zspec}{z_{\mathrm{spec}}}
\newcommand{\fmask}{f_{\mathrm{mask}}}

\newcommand{\zmax}{z_{\rm{max}}}

\shortauthors{Rykoff et al.}
\shorttitle{\redmapper{} on DES SVA1}

\begin{document}
\title{The redMaPPer Galaxy Cluster Catalog From DES Science Verification Data}

\author{
E.~S.~Rykoff\altaffilmark{1,2,$\star$},
E.~Rozo\altaffilmark{3},
D.~Hollowood\altaffilmark{4},
A.~Bermeo-Hernandez\altaffilmark{5},
T.~Jeltema\altaffilmark{4},
J.~Mayers\altaffilmark{5},
A.~K.~Romer\altaffilmark{5},
P.~Rooney\altaffilmark{5},
A.~Saro\altaffilmark{6},
C.~Vergara Cervantes\altaffilmark{5},
R.~H.~Wechsler\altaffilmark{7,1,2},
H.~Wilcox\altaffilmark{8},
T. M. C.~Abbott\altaffilmark{9},
F.~B.~Abdalla\altaffilmark{10,11},
S.~Allam\altaffilmark{12},
J.~Annis\altaffilmark{12},
A.~Benoit-L{\'e}vy\altaffilmark{13,10,14},
G.~M.~Bernstein\altaffilmark{15},
E.~Bertin\altaffilmark{13,14},
D.~Brooks\altaffilmark{10},
D.~L.~Burke\altaffilmark{1,2},
D.~Capozzi\altaffilmark{8},
A.~Carnero~Rosell\altaffilmark{16,17},
M.~Carrasco~Kind\altaffilmark{18,19},
F.~J.~Castander\altaffilmark{20},
M.~Childress\altaffilmark{21,22},
C.~A.~Collins\altaffilmark{23},
C.~E.~Cunha\altaffilmark{1},
C.~B.~D'Andrea\altaffilmark{8,24},
L.~N.~da Costa\altaffilmark{16,17},
T.~M.~Davis\altaffilmark{25},
S.~Desai\altaffilmark{26,6},
H.~T.~Diehl\altaffilmark{12},
J.~P.~Dietrich\altaffilmark{26,6},
P.~Doel\altaffilmark{10},
A.~E.~Evrard\altaffilmark{27,28},
D.~A.~Finley\altaffilmark{12},
B.~Flaugher\altaffilmark{12},
P.~Fosalba\altaffilmark{20},
J.~Frieman\altaffilmark{12,29},
K.~Glazebrook\altaffilmark{30},
D.~A.~Goldstein\altaffilmark{31,32},
D.~Gruen\altaffilmark{1,33,2,34},
R.~A.~Gruendl\altaffilmark{18,19},
G.~Gutierrez\altaffilmark{12},
M.~Hilton\altaffilmark{35},
K.~Honscheid\altaffilmark{36,37},
B.~Hoyle\altaffilmark{34},
D.~J.~James\altaffilmark{9},
S.~T.~Kay\altaffilmark{38},
K.~Kuehn\altaffilmark{39},
N.~Kuropatkin\altaffilmark{12},
O.~Lahav\altaffilmark{10},
G.~F.~Lewis\altaffilmark{40},
C.~Lidman\altaffilmark{39},
M.~Lima\altaffilmark{41,16},
M.~A.~G.~Maia\altaffilmark{16,17},
R.~G.~Mann\altaffilmark{42},
J.~L.~Marshall\altaffilmark{43},
P.~Martini\altaffilmark{36,44},
P.~Melchior\altaffilmark{45},
C.~J.~Miller\altaffilmark{27,28},
R.~Miquel\altaffilmark{46,47},
J.~J.~Mohr\altaffilmark{26,6,33},
R.~C.~Nichol\altaffilmark{8},
B.~Nord\altaffilmark{12},
R.~Ogando\altaffilmark{16,17},
A.~A.~Plazas\altaffilmark{48},
K.~Reil\altaffilmark{2},
M.~Sahl{\'e}n\altaffilmark{49},
E.~Sanchez\altaffilmark{50},
B.~Santiago\altaffilmark{51,16},
V.~Scarpine\altaffilmark{12},
M.~Schubnell\altaffilmark{28},
I.~Sevilla-Noarbe\altaffilmark{50,18},
R.~C.~Smith\altaffilmark{9},
M.~Soares-Santos\altaffilmark{12},
F.~Sobreira\altaffilmark{12,16},
J.~P.~Stott\altaffilmark{49},
E.~Suchyta\altaffilmark{15},
M.~E.~C.~Swanson\altaffilmark{19},
G.~Tarle\altaffilmark{28},
D.~Thomas\altaffilmark{8},
D.~Tucker\altaffilmark{12},
S.~Uddin\altaffilmark{30},
P.~T.~P.~Viana\altaffilmark{52,53},
V.~Vikram\altaffilmark{54},
A.~R.~Walker\altaffilmark{9},
Y.~Zhang\altaffilmark{28}
\\ \vspace{0.2cm} (The DES Collaboration) \\
}
\email{$\star$ erykoff@slac.stanford.edu}

\begin{abstract}
  We describe updates to the \redmapper{} algorithm, a photometric red-sequence
  cluster finder specifically designed for large photometric surveys.  The
  updated algorithm is applied to $150\,\mathrm{deg}^2$ of Science Verification
  (SV) data from the Dark Energy Survey (DES), and to the Sloan Digital Sky
  Survey (SDSS) DR8 photometric data set.  The DES SV catalog is locally volume
  limited, and contains 786 clusters with richness $\lambda>20$ (roughly
  equivalent to $M_{\rm{500c}}\gtrsim10^{14}\,h_{70}^{-1}\,M_{\sun}$) and
  $0.2<z<0.9$.  The DR8 catalog consists of 26311 clusters with $0.08<z<0.6$,
  with a sharply increasing richness threshold as a function of redshift for
  $z\gtrsim 0.35$.  The photometric redshift performance of both catalogs is
  shown to be excellent, with photometric redshift uncertainties controlled at
  the $\sigma_z/(1+z)\sim 0.01$ level for $z\lesssim0.7$, rising to $\sim0.02$
  at $z\sim0.9$ in DES SV.  We make use of \emph{Chandra} and \emph{XMM} X-ray
  and South Pole Telescope Sunyaev-Zeldovich data to show that the centering
  performance and mass--richness scatter are consistent with expectations based
  on prior runs of \redmapper{} on SDSS data.  We also show how the
  \redmapper{} \photoz{} and richness estimates are relatively insensitive to
  imperfect star/galaxy separation and small-scale star masks.
\end{abstract}


\section{Introduction}

Clusters of galaxies are the largest bound objects in the Universe, and are
uniquely powerful cosmological probes~\citep[e.g.,][]{henryetal09,
  vikhlininetal09b, mantzetal10a, rwrab10, clercetal12, bensonetal13,
  hasselfieldetal13, planckXX14} (see also reviews in
\citet{allenetal11,weinbergetal13}).  In particular, galaxy clusters are one of
the key probes of growth of structure and dark energy measurements from ongoing and
upcoming photometric surveys such as the Dark Energy
Survey~\citep[DES;][]{des05}, the Kilo-Degree Survey~\citep[KiDS;][]{kidsdr1},
the Hyper-Suprime Camera
(HSC)\footnote{http://www.naoj.org/Projects/HSC/HSCProject.html}, the Large
Synoptic Survey Telescope~\citep[LSST;][]{lsst09},
\emph{Euclid}~\citep{euclid11}, and
WFIRST\footnote{http://wfirst.gsfc.nasa.gov/}.

There are already a wide range of photometric cluster
finders~\citep[e.g.][]{gotoetal02, gladdersetal07, koesteretal07a, hmkrr10,
  soaressantosetal11, spdpg11, whl12, mgb12, ascasoetal12, ascasoetal14,
  oguri14}, each with various strengths and weaknesses.  In 2014, we introduced
the {\bf red}-sequence {\bf ma}tched-filter {\bf P}robabalistic {\bf
  Per}colation cluster finder~\citep[\redmapper;][henceforth RM1]{RM1}.
\Redmapper{} identified galaxy clusters by making use of the fact that the bulk
of the cluster population is made up of old, red galaxies with a prominent
$4000\,\mathrm{\AA}$ break.  Focusing on this specific galaxy population
increases the contrast between cluster and background galaxies in color space,
and enables accurate and precise photometric redshift (\photoz) estimates.  The
associated cluster richness estimator, $\lambda$, is the sum of of the
membership probability of every galaxy in the cluster field, and has been
optimized to reduce the scatter in the richness--mass relation~\citep{rrkmh09,
  rrknw11, rykoffetal12}.

The initial application of \redmapper{} in RM1 was on the Sloan Digital Sky
Survey Data Release 8 photometric data~\citep[SDSS DR8;][]{yorketal00,dr8}.  As
such, the catalog was limited to relatively low redshifts ($z\lesssim0.5$).
The SDSS \redmapper{} catalog has been extensively validated using
X-ray~\citep[][henceforth RM2]{RM2}; \citep{sadibekovaetal14} and
Sunyaev-Zeldovich (SZ) data~\cite[][henceforth RM3]{RM3}, and with
spectroscopic data~\citep[][henceforth RM4]{RM4}, demonstrating that the
catalog has low scatter in the mass--richness relation; well-quantified
centering performance; and accurate and precise cluster \photozs.  The low
scatter has also made it possible to use the \redmapper{} SDSS catalog to
verify Planck clusters~\citep{RM3, planckXXVII15}.  In a comparison of numerous
spectroscopic cluster finders on mock catalogs, \redmapper{} achieved one of
the smallest variances in estimated cluster mass at fixed halo mass, despite
being the only cluster finder relying solely on two-band photometric data (all
other cluster finders were spectroscopic)~\citep{oldetal15}.

\Redmapper{} was designed to easily handle a broad range in redshift, as well
as to run efficiently over a wide and deep galaxy catalog.  As such, it is
ideally suited to DES data, which can be used to detect faint red-sequence galaxies to
much higher redshifts than SDSS ($z\lesssim0.9$).  In this paper, we describe
the first application of \redmapper{} to DES Science Verification (SV) data.
In addition, we describe updates to the \redmapper{} algorithm since versions
5.2 (RM1) and 5.10 (RM4) to the present version 6.3, and apply the updated
algorithm to the SDSS DR8 photometric data.  
We characterize the \photoz\ performance of \redmapper\ using available
spectroscopy, and use available SZ data from the 
South Pole Telescope SZ cluster
survey~\citep[SPT;][]{bleemetal15}, as well as 
X-ray observations from \emph{Chandra} and \emph{XMM},
to measure the centering
properties of the DES SV \redmapper{} catalog, and to test
the validity of the \redmapper\ cluster richness as a photometric mass tracer.
A detailed analysis of the richness and SZ scaling relations is
presented in \citet[][henceforth S15]{saroetal15}.  A similar analyis
of X-ray observations including SDSS overlap will be presented in 
Bermeo Hernendez et al. (in prep) and Hollowood et al. (in prep).

The layout of this paper is as follows.  In Section~\ref{sec:data} we describe
the DES Science Verification and SDSS DR8 data used in this work.
Section~\ref{sec:updates} describes the updates to the \redmapper{} algorithm
since the RM1 and RM4 papers.  Section~\ref{sec:catalog} describes the cluster
catalogs, as well as the photometric redshift performance on DES and SDSS
data.  In Section~\ref{sec:sgmask} we detail the effects of star/galaxy
separation and small-scale masking on the cluster properties, and in
Section~\ref{sec:xraysz} we compare the \redmapper{} catalog with X-ray and SZ
clusters in the DES SVA1 footprint.  Finally, in Section~\ref{sec:summary} we
summarize our results.  When necessary, distances are estimated assuming a flat
$\Lambda\mathrm{CDM}$ model with $\Omega_m = 0.30$.  For consistency with
previous \redmapper{} work, we use $h=1.0$ when quoting distances
($h^{-1}\,\mathrm{Mpc}$) and $h=0.7$ when quoting masses ($h_{70}^{-1}\,\msun$).

\section{Data}
\label{sec:data}

\subsection{DES Science Verification Data}
\label{sec:sva1data}

The DES is an ongoing 5-band ($grizY$) photometric survey performed with the
Dark Energy Camera~\citep[DECam,][]{decam15} on the 4-meter Blanco
Telescope at Cerro Tololo Inter-American Observatory (CTIO).  Prior to the
beginning of the DES survey, from November 2012 to March 2013, DES conducted a
$\sim250\,\mathrm{deg}^2$ ``Science Verification'' (SV) survey.  The largest
contiguous region covers $\sim 160\,\mathrm{deg}^2$ of the eastern edge of the
SPT survey (``SPT-E'' hereafter).  A smaller $\sim 35\,\mathrm{deg}^2$ region
is in the western edge of the footprint (``SPT-W'' hereafter).  In addition,
DES surveys 10 Supernova fields (``SN fields'' hereafter) every 5-7 days, each
of which covers a single DECam 2.2-degree-wide field of view, for a total of
$\sim32\,\mathrm{deg}^2$ of deeper imaging (including extra offset pointings of
SN fields taken during SV).  Finally, there are smaller
discontinguous regions targeting massive clusters~\citep{melchioretal15} and
the COSMOS field~\citep{cosmos07}.  We utilize this DES SV data set to
construct the first DES \redmapper{} cluster catalog.  The \redmapper{}
footprint used in this paper is the same as that used for the associated
\redmagic{} ({\bf red}-sequence {\bf ma}tched-filter {\bf G}alaxies {\bf
  Catalog}) of red galaxies with well-behaved \photoz{}
performance~\citep[][henceforth RM15]{redmagic15}.

The DES SV data was processed by the DES Data Management (DESDM)
infrastructure~(Gruendl et al, in prep), which includes 
image detrending, astrometric registration, global calibration, image
coaddition, and object catalog creation.  Details of the DES single-epoch and
coadd processing can be found in \citet{sevillaetal11} and
\citet{desaietal12}.  We use {\tt SExtractor} to create object catalogs from
the single-epoch and coadded images~\citep{bertinarnouts96, bertin11}.  

After the initial production of these early data products, we detected several
issues that were mitigated in post-processing, leading to the creation of the
``SVA1 Gold'' photometry
catalog\footnote{http://des.ncsa.illinois.edu/releases/sva1}.  First, we masked
previously unmasked satellite trails.  Second, we use a modified version of the
{\tt big-macs} stellar-locus regression (SLR) fitting
code~\citep{kellyetal14}\footnote{https://code.google.com/p/big-macs-calibrate/}
to recompute coadd zero-points over the full SVA1 footprint.  Third, regions
around bright stars ($J<13$) from the Two Micron All Sky
Survey~\citep[2MASS;][]{skrutskieetal06} were masked.  Finally, we removed
$4\%$ of the area with a large concentration of centroid shifts between
bandpasses in individual objects, indicating scattered light, ghosts, satellite
trails, and other artifacts~\citep[][Section 2.1]{jarvisetal15}.  We utilize
the {\tt SExtractor} \magauto{} quantity derived from the coadded images for
galaxy total magnitudes and colors.  This choices reflects the fact that
methods to compute multi-epoch fitting photometric quantities are still under
development.  The added noise in the color results in a larger observed
red-sequence width, which results in slightly poorer photometric redshifts, as
shown in Section~\ref{sec:sva1photozperf}. For the present work, we have not
made use of the DES $Y$-band imaging because of uncertain calibration, and the
minimal lever-arm gained at the redshifts probed in this paper.  Finally, our
fiducial star/galaxy separation is done using the multi-band multi-epoch image
processing code \ngmix{} used for galaxy shape measurement in DES
data~\citep{jarvisetal15}, as detailed in Appendix A of RM15.

\begin{figure}
  \plotone{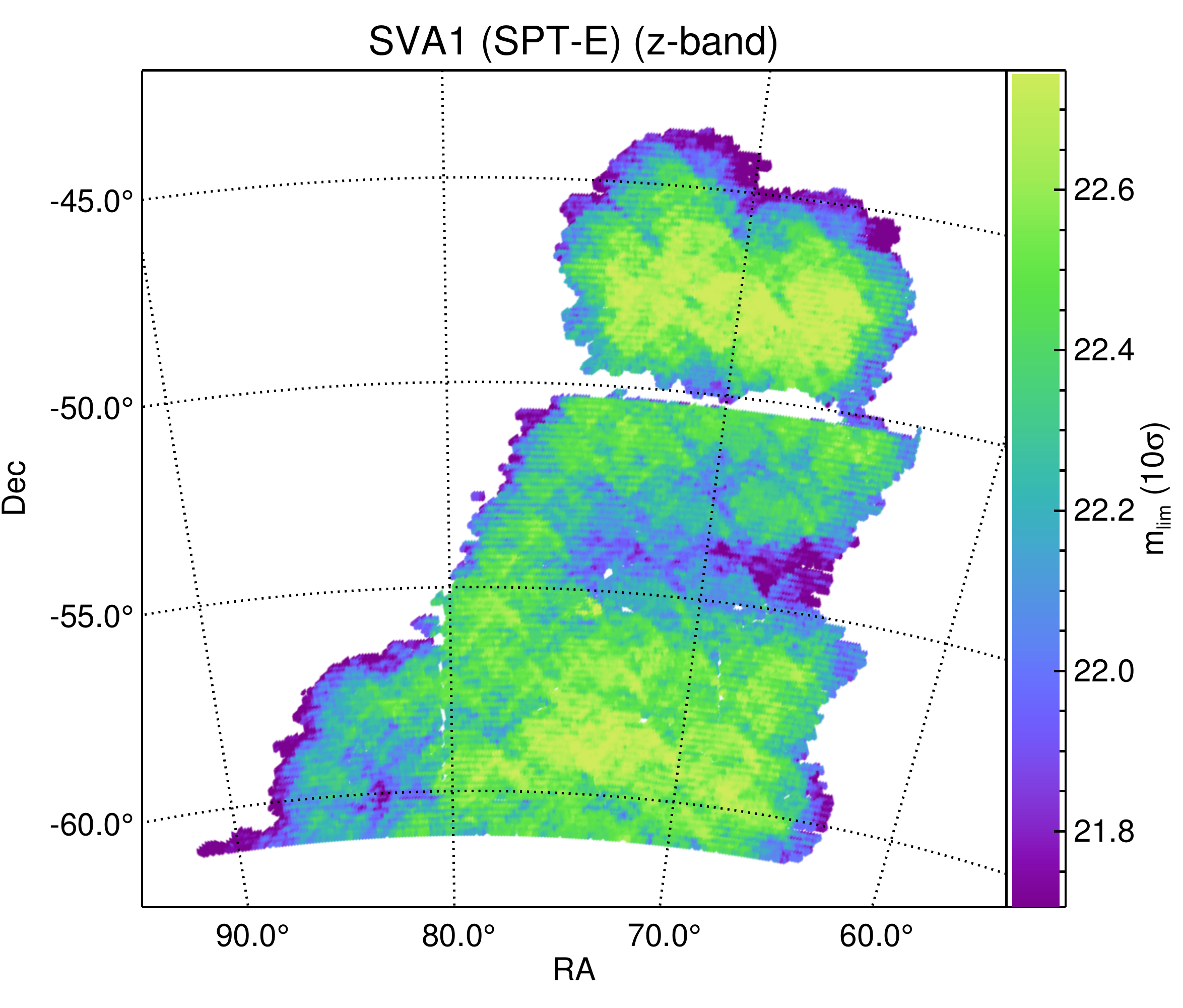}
  \caption{Map of $10\sigma$ depth (in magnitudes) in $\sim125\,\mathrm{deg}^2$
    in the SVA1 SPT-E footprint, for
    SLR-corrected $z_{\mathrm{auto}}$ magnitudes.  Small scale variations are
    caused by variations in the number of exposures, chip gaps, and observing conditions.}
  \label{fig:sva1depth}
\end{figure}

The footprint is initially defined by \mangle{}~\citep{sthh08} maps generated
by DESDM which describe the geometry of the coadded data in polygons of
arbitrary resolution.  For ease of use, these are then averaged over \healpix{}
\nsidef{} pixels~\citep{gorskietal05}, where each pixel is approximately $0.7'$
on a side.  The pixelized \mangle{} maps are combined with maps of the survey
observing properties (e.g., airmass, full-width-half-maximum, etc.) compiled by
\citet{leistedtetal15} using the method of \citet{rykoffdepth15} to generate
$10\sigma$ \magauto{} limiting magnitude maps.  We first restrict the footprint to
regions with deep {\tt MAG\_AUTO} on the $z$ band ($\mzlim$) such that $\mzlim > 22$,
as shown in Figure~\ref{fig:sva1depth} for $\sim125\,\mathrm{deg}^2$ in the
SPT-E region.\footnote{This is equivalent to a
  $0.2L*$ galaxy at $z=0.65$, as described in Section~\ref{sec:mstar}} Only
galaxies brighter than the local $10\sigma$ limiting magnitude are used in the
input catalog.

The \ngmix{} runs used for star/galaxy separation in this paper and RM15 were
primarily used for galaxy shape estimation for DES cosmic
shear~\citep{beckeretal15} and cosmological constraints~\citep{desetal15}.
Therefore, the runs were performed on regions with very tight tolerance for
image quality, and were restricted to the largest contiguous region (SPT-E) as
well as supplementary runs on the SN fields.  These regions comprise our
fiducial footprint for the input galaxy catalog of $148\,\mathrm{deg}^2$ (of
which $125\,\mathrm{deg}^2$ is in SPT-E).  However, mask boundaries and
holes reduce the effective area for extended cluster sources to $\sim100\,\mathrm{deg}^2$ (see
Section~\ref{sec:randpts} for details).  In Section~\ref{sec:sgmask} we
describe an expanded footprint where we relax some of these constraints, and
include SPT-W and COSMOS, with less robust star/galaxy separation.

Spectroscopic data used in this paper comes from the Galaxy and Mass
Assembly survey~\citep[GAMA][]{gamadr1}, the VIMOS VLT Deep
Survey~\citep[VVDS][]{garillietal08}, the 2dF Galaxy Redshift
Survey~\citep[2dFGRS][]{collessetal01}, the Sloan Digital Sky
Survey~\citep[SDSS][]{dr10}, the VIMOS Public Extragalactic
Survey~\citep[VIMOS][]{garillietal14}, and the Arizona CDFS Environment
Survey~\citep[ACES][]{cooperetal12}. In addition, we have a small sample of
cluster redshifts from SPT used in the cluster
validation of ~\citet{bleemetal15}. These data sets have been further supplemented
by galaxy spectra acquired as part of the OzDES spectroscopic survey, which is
performing spectroscopic follow-up on the AAOmega instrument at the
Anglo-Australian Telescope (AAT) in the DES supernova fields \citep{ozdes15}.
In all, there are 36,607 photometric galaxies with spectroscopic redshifts in
our input catalog, although only $\sim2000$ are red cluster members, and
$\sim1400$ are used in the calibration of the red sequence in
Section~\ref{sec:initselect}.

\subsection{SDSS DR8}

In addition to our new catalog on DES SVA1 data, we have updated the
\redmapper{} catalog for SDSS DR8 photometric
data~\citep{dr8}, which remains the most recent photometric data release from SDSS.
The DR8 galaxy catalog contains $\sim 14000\,\mathrm{deg}^2$ of imaging, which
we cut to the $10401\,\mathrm{deg}^2$ of contiguous high quality observations
using the mask from the Baryon Acoustic Oscillation
Survey~\citep[BOSS,][]{dsaaa13}.  The mask is further extended to exclude all
stars in the Yale Bright Star Catalog~\citep{hoffleitjaschek91}, as well as the
area around objects in the New General Catalog \citep[NGC][]{sinnott88}.  The
resulting mask is that used by RM1 to generate the SDSS DR8
\redmapper\ catalog.  We refer the reader to that work for further discussion
on the mask, as well as object and flag selection.

Total magnitudes are determined from $i$-band SDSS {\tt CMODEL\_MAG}, which we
denote $\mi$, and colors from $ugriz$ SDSS {\tt MODEL\_MAG}.  All spectroscopy
is drawn from SDSS DR10~\citep{dr10}.  Finally, we make use of the $10\sigma$
limiting magnitude maps from~\citet[][see, e.g., Figure 4]{rykoffdepth15}.  As
with SVA1 data, only galaxies brighter than the local $10\sigma$ limiting
magnitude are used in the input catalog.  

\section{Updates to the \redmapper{} Algorithm}
\label{sec:updates}

\label{sec:redmapperintro}

\Redmapper{} is a matched-filter red-sequence photometric cluster finding
algorithm with three filters based on galaxy color, position, and luminosity.
The most important filter characterizes the color of red-sequence galaxies as a
function of redshift.  This filter is a linear red-sequence model in
color-magnitude space (with slope and intercept) in $n_{\mathrm{col}}$
dimensions, where $n_{\mathrm{col}}$ is the number of independent colors in the
input dataset.  The filter also incorporates the intrinsic scatter,
$C_{\mathrm{int}}$, which is the $n_{\mathrm{col}} \times n_{\mathrm{col}}$
covariance matrix assuming Gaussian errors in photometric magnitudes.  This
filter is self-calibrated by making use of clusters with known spectroscopic
redshifts.  The additional two filters are the radial filter, comprised of a
projected NFW profile~\citep{navarro_etal94}, and a luminosity filter based on
a Schechter function.  Once the parameters of the red-sequence filter is known, we use this
information to compute a probability $\pmem$ that each galaxy in the vicinity
of the cluster is a red-sequence member.  The richness $\lambda$ is defined as
the sum of the membership probabilities over all galaxies within a scale-radius
$R_\lambda$:
\be
\lambda = \sum \pmem \theta_L \theta_R,
\ee
where $\theta_L$ and $\theta_R$ are luminosity- and radius-dependent weights
defined in Appendix B of RM4.  The radius scales with the size of the cluster
such that $R_\lambda = 1.0 (\lambda/100)^{0.2} \hMpc$, which we have shown
minimizes the scatter in the mass--richness relation~\citep{rkrae12}.  All
galaxies with magnitudes consistent with being brighter than $0.2\,L_*$ are
considered for computing the richness, as described below in
Section~\ref{sec:mstar}.  We note that the weights $\theta_L$ and $\theta_R$
are ``soft cuts'' to ensure cluster richness measurements are robust
to small perturbations in galaxy magnitudes.
The cluster photometric redshift, $\zlambda$, is constrained at the same time
as the cluster richness by fitting all possible member galaxies simultaneously
to the red-sequence color function.

The above equation describes the richness computation in the absence of any
masking (star holes and survey boundaries), and in the regime where the local
limiting magnitude is deeper than $0.2\,L_*$ at the cluster redshift. As
described in Section 5 of RM1, we additionally compute a scale factor $S$ to
correct for these missing cluster members, such that
\be
\frac{\lambda}{S} =  \sum_{\mathrm{gals}} \pmem
\label{eqn:S}
\ee
so that each cluster with richness $\lambda$ has $\lambda/S$ galaxies brighter
than the limiting magnitude of the survey within the geometric survey mask.  At
the same time, we estimate the variance $S$ which is used in the computation of
the uncertainty on richness $\lambda$, as detailed in Appendix B of RM4.  In this way, the
total uncertainty on $\lambda$ includes the uncertainty from correcting for mask and
depth effects.

In addition, as described in Section 5.1 of RM1 (specifically Eqn. 24), it is
useful to compute the fraction of the effective cluster area that is masked
solely by geometric factors such as stars, bad regions, and survey edges.  This
mask fraction, denoted $\fmask$, is complementary to $S$ above in that it
contains all the local masking except the depth limit.

As well as estimating membership probabilities, the \redmapper{} centering
algorithm is also probabilistic (see Section 8 of RM1).  The centering
probability $\Pcen$ is a likelihood-based estimate of the probability that a
galaxy under consideration is a central galaxy (CG).  This likelihood includes
the fact that the \photoz{} of the CG must be consistent with the cluster
redshift; that the CG luminosity must be consistent (using a Gaussian filter) with the expected
luminosity of the CG of a cluster of the observed richness; and that the local
red galaxy density (on the scale of $\sim200\,h^{-1}\,\mathrm{kpc}$) is
consistent with that of CGs.  We additionally assume that each cluster can have
at most one CG, and store the top 5 most likely central candidates.  Our
fiducial cluster position is given by the highest likelihood central galaxy.  Because of the luminosity filter, the CG
candidate with the largest $\Pcen$ tends to be very bright, but is not
necessarily the brightest member. Thus, we do not refer to it as the
brightest cluster galaxy (BCG), only as the central galaxy.  
Typically, for $\sim15-20\%$ of the clusters
the CG chosen by \redmapper{} is not the brightest member.

The \redmapper{} algorithm has previously been applied to SDSS DR8
photometric data.  For more details on the \redmapper{} algorithm, we refer
the reader to RM1 and the updates in the appendix of RM4.  In this section, we
detail the various modifications that have been implemented
on the \redmapper\ algorithm since its last public data release (RM4).

\subsection{Incorporating Small Scale Structure in the Local Survey Depth}
\label{sec:depthmaps}

Variable survey depth can lead to galaxies being ``masked out'' from galaxy
clusters.  Specifically, if a member galaxy (with $L\geq0.2L_*$) has a
magnitude below our brightness threshold, then one needs to
statistically account for this missing galaxy, as per the above formalism.  To
do so, however, one needs to know the survey depth over the full area coverage of
the galaxy cluster.

The original \redmapper{} application to SDSS DR8 in RM1 (\redmapper{} v5.2)
assumed the survey had a uniform depth with $\mi<21.0$.  In the update
described in RM4 (\redmapper{} v5.10), we empirically computed the local survey
depth averaged over the location of each cluster.  This was superior to
assuming a constant-depth survey, but ignored small-scale depth variations, as
well as being somewhat noisy.  In this updated version version (\redmapper{}
v6.3), we have extended \redmapper{} to incorporate variable survey limiting
magnitude maps as detailed in \citep{rykoffdepth15} and described in
Section~\ref{sec:data}.  Specifically, we utilize the local survey depth from
these depth maps to estimate the fraction of cluster galaxies that are
masked, as defined in Section~\ref{sec:redmapperintro} and detailed in Appendix
B of RM4.  In the present version of the algorithm, we assume that the red
galaxy detection is complete (modulo masking) at magnitudes brighter than the
local $10\sigma$ limiting magnitude used to select the input catalog.  In
future versions, we intend to track the full completeness function, as
described in Section~5 of \citet{rykoffdepth15}.  

\subsection{Generalization of the Characteristic Magnitude $m_*(z)$ to Arbitrary
Survey Filters}
\label{sec:mstar}

As with the previous versions of the \redmapper{} algorithm our luminosity
filter is based on a Schechter function~\citep[e.g.,][]{hswk09} of the form:
\begin{equation}
\phi(\mi) \propto 10^{-0.4(\mi-m_*)(\alpha+1)}\exp\left(-10^{-0.4(\mi-m_*)}\right),
\label{eqn:lumfilter}
\end{equation}
where we set the faint-end slope $\alpha=1.0$ independent of redshift.
Previously, we set the characteristic magnitude, $m_*(z)$, using a
$k$-corrected passively evolving stellar population which we had derived from a
PEGASE.2 stellar population/galaxy formation model~\citep{pegase,kmawe07a,
  eisensteinetal01}.  As this was derived specifically for the SDSS filters at
relatively low redshift, we have updated our reference $m_*(z)$ to more simply
allow for different filter sets and a broader redshift range.

The new value of $m_*(z)$ is computed using a \citet[][BC03]{bruzualcharlot03}
model to predict the magnitude evolution of a galaxy with a single star
formation burst at $z=3$ (with solar metallicity and Salpeter IMF) as implemented in the {\tt EzGal} Python
package~\citep{ezgal12}.  We normalize $m_*$ so that $m_{i,\mathrm{SDSS}} =
17.85$ at $z=0.2$ for an $L_*$ galaxy. This was chosen to match the $m_*(z)$
relation from RM1 and \citet{rykoffetal12}.  We have additionally confirmed
that the evolution of $m_*(z)$ is within $8\%$ of that used in RM1 over the RM1
redshift domain ($0.1<z<0.5$), with the largest deviations at $z\sim0.5$.  The
normalization condition for $\mz$ for DES is then derived from the BC03 model
using the DECam passbands~\citep{decam15}.

\subsection{Initial Selection of Red Spectroscopic Galaxies}
\label{sec:initselect}

As described in RM1, the initial calibration of the red sequence relies on
spectroscopic ``seed'' galaxies.  This may be comprised of a set of training
clusters with spectroscopic redshifts (as in DES SVA1) or a large spectroscopic
catalog with a sufficient number of red galaxies in clusters (as in SDSS DR8).
In RM1 (see Section 6.2), we selected red galaxies by splitting the
spectroscopic catalog into narrow redshift bins, and using a Gaussian mixture
model~\citep{haoetal09} to separate galaxies in each redshift bin into blue and
red components.  However, we have found that this method is only robust when we
have a plethora of spectra, as is the case with SDSS.

In this paper, the initial red galaxy selection is performed by computing color
residuals of galaxies in a broad range of redshifts relative to the
BC03-derived color models from Section~\ref{sec:mstar}.  As we are only
concerned with making an initial selection of red and blue galaxies, any color
calibration offsets between the data and the BC03 model are irrelevant; we just
need to get an initial sample of red galaxies.  We again employ a Gaussian
mixture model to obtain a first estimate for the mean color and intrinsic
scatter of the red spectroscopic galaxies.  To ensure a clean selection, we use
the $g-r$ color for $\zspec<0.35$; $r-i$ for $0.35<\zspec<0.7$; and $i-z$ for
$\zspec>0.7$.  At this point, we proceed as described in Step 3 of RM1 Section
6.2.

\subsection{Redshift Reach of the Cluster Catalog}
\label{sec:vlim}

Ideally, a photometric survey would be deep enough to detect the faintest $0.2L_*$
galaxies that contribute to our richness estimator $\lambda$ over the full
redshift range and footprint of the catalog.  In a roughly uniform survey such
as the SDSS, this limitation translates into a maximum redshift, $\zmax$, below
which the cluster catalog is volume limited; for SDSS, $\zmax<0.33$.   By contrast, 
the observing strategy of a multi-epoch survey such as the DES may yield much greater
depth variations, as shown in Figure~\ref{fig:sva1depth}.  Furthermore, the
depth variations can be different in different bands.  Consequently, the redshift
range which can be successfully probed with \redmapper\ will depend on
the local survey depth, with deeper regions allowing us to detect galaxy
clusters to higher redshifts.

We define a maximum redshift $\zmax$ at each position in the sky as follows.
Given a point in the survey, our initial depth map for the main detection band
($\mz$ in the case of SVA1), and a luminosity threshold
($\Lthresh$), we calculate the maximum redshift to which a typical
red galaxy (defined by our red-sequence model) of $\Lthresh$ is:
detectable at $>10\sigma$ in the main detection band ($z$-band for DES); and at
$>5\sigma$ in the remaining bands.  Only clusters with $z\leq \zmax$ are
accepted into our cluster catalog, with $\zmax$ defining the redshift component
of our survey mask.  In this way, we can simply (and conservatively) account
for the regions that are extremely shallow in one or more bands.  This happens
in SVA1 primarily at the boundaries, and other regions that were observed in
non-photometric conditions.  The result is a map of $\zmax$ in \healpix{} format
with \nsidef, where each pixel is approximately $0.7'$
on a side, that is matched to the resolution of the input depth maps.

Given this procedure, we still have an arbitrary decision as to where to set
our luminosity threshold $\Lthresh$.  The most conservative option would be to
demand that every cluster in the final catalog be at a redshift such that we
can detect red galaxies to $\Lthresh = 0.2L_*$.  However, we have chosen to be
somewhat more aggressive in the interest of increasing the number of galaxy
clusters and redshift reach of the DES SV catalog, as the impact on the uncertainty
estimate of $\lambda$ (see Eqn.~\ref{eqn:S}) is modest for clusters that only
require a small extrapolation.  For SVA1, we have chosen the luminosity
threshold to be $\Lthresh = 0.4L_*$ for the construction of the $\zmax$ map.
For DR8, we have chosen $\Lthresh = 1.0L_*$, such that $\zmax\sim0.6$ over
$>99\%$ of the DR8 footprint.  Although this requires a large richness
extrapolation at high redshift (and hence large richness errors), this cut
maintains consistency with previous \redmapper{} catalogs (versions 5.2 and
5.10) where we did not use a $\zmax$ map.  However, if users wish to utilize a
volume-limited subset of the DR8 \redmapper{} catalog, restricting to
$\zlambda<0.33$ will ensure that the local depth at every cluster is deep
enough to detect $0.2L_*$ galaxies.

\subsection{Differences Between the SVA1 and DR8 Analyses}

Although the code used to run on SVA1 and DR8 is the same, there are a few key
differences that we highlight here.

\begin{enumerate}
  \item{For DR8, we use $i$-band for the detection magnitude; for SVA1 we use $z$-band,
    which is better suited to the broad redshift range and the excellent
    $z$-band performance of DECam.}
  \item{For DR8, we use $ugriz$ for galaxy colors, while for SVA1 we only use
    $griz$.  The lack of $u$ band has negligible effect on the cluster detection
    and cluster \photozs{} at $z>0.2$ (see Section 8.1 of RM15).}
  \item{For DR8, reddening corrections are applied to catalog magnitudes.  For
    SVA1, these are incorporated into the SLR zero-point calibration.}
  \item{For DR8, we train the red sequence model over $2000\,\mathrm{deg^2}$
    ($\sim20\%$ of the full footprint), as in RM1, to ensure sufficient
    statistics of spectroscopic training while avoiding any possibility of
    over-training.  For the much smaller SVA1 catalog, we use the full
    footprint and all available spectra.  The impact of this is detailed in
    Section~\ref{sec:sva1photozperf}.}
\end{enumerate}

\subsection{Generation of Random Points}
\label{sec:randpts}

In RM1, we describe a method of estimating purity and completeness of the
cluster catalog using the data itself, by placing fake clusters into the data
and recovering the richness.  While this method (described in Section 11 of
RM1) is useful for estimating the selection function and projection effects, it
is not appropriate for generating a cluster random catalog for
cross-correlation measurements, such as the cluster--shear cross-correlation
used for stacked weak-lensing mass estimates~\citep[e.g.,][]{johnstonetal07, reyesetal08}, as
existing large-scale structure is imprinted on the random catalog.  

In this section, we describe a new way of generating cluster random points by
making use of the $\zmax$ map from Section~\ref{sec:vlim}.  A particular
challenge is the fact that galaxy clusters are extended objects, and thus the
detectability depends not just on the redshift, but the cluster size and the
survey boundaries. We generate a random cluster catalog that has the same
richness and redshift distribution of the data catalog by randomly sampling
$\{\lambda,\zlambda\}$ pairs from the data catalog.  To ensure that the random
catalog correctly samples the survey volume, we utilize the redshift mask.
Specifically, after sampling a cluster from the cluster catalog, we randomly
sample a position ($\{\alpha, \delta\}$) for the random point.  If the cluster
redshift $\zlambda$ is larger than the maximum redshift at which the cluster
can be detected, we draw a new $\{\alpha, \delta\}$, repeating the procedure
until the cluster is assigned a position consistent with the cluster
properties.  In all, we sample each cluster $\nsamp\sim1000$ times to ensure
that any correlation measurements we make are not affected by noise in the
random catalog.

Having assigned a position, we use the depth map and the footprint mask to
estimate the local mask fraction $\fmask$ and scale factor $S$, as defined in
Section~\ref{sec:redmapperintro}.  This is the point at which the finite extent
of the clusters is taken into account.  Only random points that have $\fmask<0.2$
and $\lambda/S>20$ are properly within the cluster detection
footprint.  These cuts will locally
modify the richness and redshift distribution of the random points relative
to the data.   In particular, the random points will tend to undersample
the regions from which we discard clusters, particularly for low richness
and high redshift clusters.

We address this difficulty by using weighted randoms.  Specifically, 
given all $\nsamp$ random points generated from a given
$\{\lambda,\zlambda\}$ pair, we calculate the number of random points that pass
our mask and threshold cuts, denoted $\nkeep$.  Each random point is then
upweighted by a factor $w = \nsamp/\nkeep$.  This ensures that the weighted
distribution of random points matches the cluster catalog as a function of both
$\lambda$ and $\zlambda$, while taking into account all boundaries and depth
variations.  As we typically sample each cluster $\sim1000$ times,
the weight $w$ is sufficiently well measured that we neglect noise in $w$ when making
use of the weighted random points.  We note that in this procedure we neglect
sample variance from large-scale structure that may be imprinted in the cluster
catalog; while this may be a small issue for SVA1, this will be averaged out
over large surveys such as DES and SDSS.

Finally, we compute the effective area of the survey for cluster detection.
For any given redshift $z$, we compute the total area ($A_{\mathrm{tot}}$)
covered where we might have a chance of detecting a cluster, such that
$z<\zmax$.  Taking into account boundaries and the finite size of clusters, the
effective area is simply $A_{\mathrm{tot}} \times \nsamp/\nkeep$, where
$\nsamp/\nkeep$ is computed for all random points with $z<\zmax$.  We then use
a cubic spline to perform a smooth interpolation as a function of redshift.
Due to the finite size of the clusters and the small footprint of SVA1 with a
lot of boundaries, the effective area for $\lambda>20$ cluster detection is
reduced from $148\,\mathrm{deg}^2$ to $\sim100\,\mathrm{deg}^2$ at $z<0.6$.


\section{The Fiducial Cluster Catalogs}
\label{sec:catalog}

We have run the updated \redmapper{} v6.3 algorithm on SDSS DR8 and DES SVA1
data as described in Section~\ref{sec:data}.  Following RM1, the full cluster
finder run contains all clusters with $\lambda \geq 5\,S$, over the redshift
range $\zlambda \in [0.05,0.6]$ (for DR8) and $\zlambda \in [0.15,0.9]$ (for
SVA1).  However, we have chosen to apply relatively conservative cuts to our
catalogs.  The cuts we apply are as follows.

\begin{enumerate}
  \item{There must be at least 20 unmasked galaxies brighter than the local limiting
    magnitude, such that $\lambda/S > 20$.}
  \item{The volume limited mask for SVA1 is as described above.  The volume limited
    catalog for DR8 is simply $\zlambda<0.33$.}
  \item{For the DR8 catalog, the richness scale factor $S(z)$ is illustrated by
    Figure 19 in RM1.  For the volume-limited SVA1 catalog, $S(z)\lesssim1.3$
    at all redshifts by construction.}
  \item{Very low redshift clusters have biased redshifts and richnesses due to
      boundary effects, so we have set the lower redshift limit $\zlambda >
      0.08$ and $\zlambda > 0.2$ for the DR8 and SVA1 catalogs, respectively.}
  \item{Only clusters with $\fmask < 0.2$ are included.  That is, clusters near
    the boundary and on top of masked regions will be removed.  The cluster
    random points properly sample the footprint, reflecting these cuts.}
\end{enumerate}
A summary of the number of clusters, effective area, and redshift range of the
catalogs (including the SVA1 expanded catalog described in Section~\ref{sec:masking})
is given in Table~\ref{tab:samples}.

\begin{deluxetable}{llll}
\tablewidth{0pt}
\tablecaption{\redmapper{} Cluster Samples}
\tablehead{
 \colhead{Sample} &
 \colhead{Area $(\mathrm{deg}^2)$}\tablenotemark{a} &
 \colhead{Redshift Range} &
 \colhead{No. of Clusters\tablenotemark{b}}
}
\startdata
DR8 & $10134$ & $0.08<\zlambda<0.6$ & 26111\\
SVA1 & $116$ & $0.2<\zlambda<0.9$ & 787\\
SVA1 expanded & $208$ & $0.2<\zlambda<0.9$ & 1382\\
\enddata
\tablenotetext{a}{Area including effect of $\fmask<0.2$ cut for extended
  cluster sources (see Section~\ref{sec:randpts}).}
\tablenotetext{b}{Richness threshold, $\lambda/S > 20$}
\label{tab:samples}
\end{deluxetable}

\begin{figure*}
  \plotone{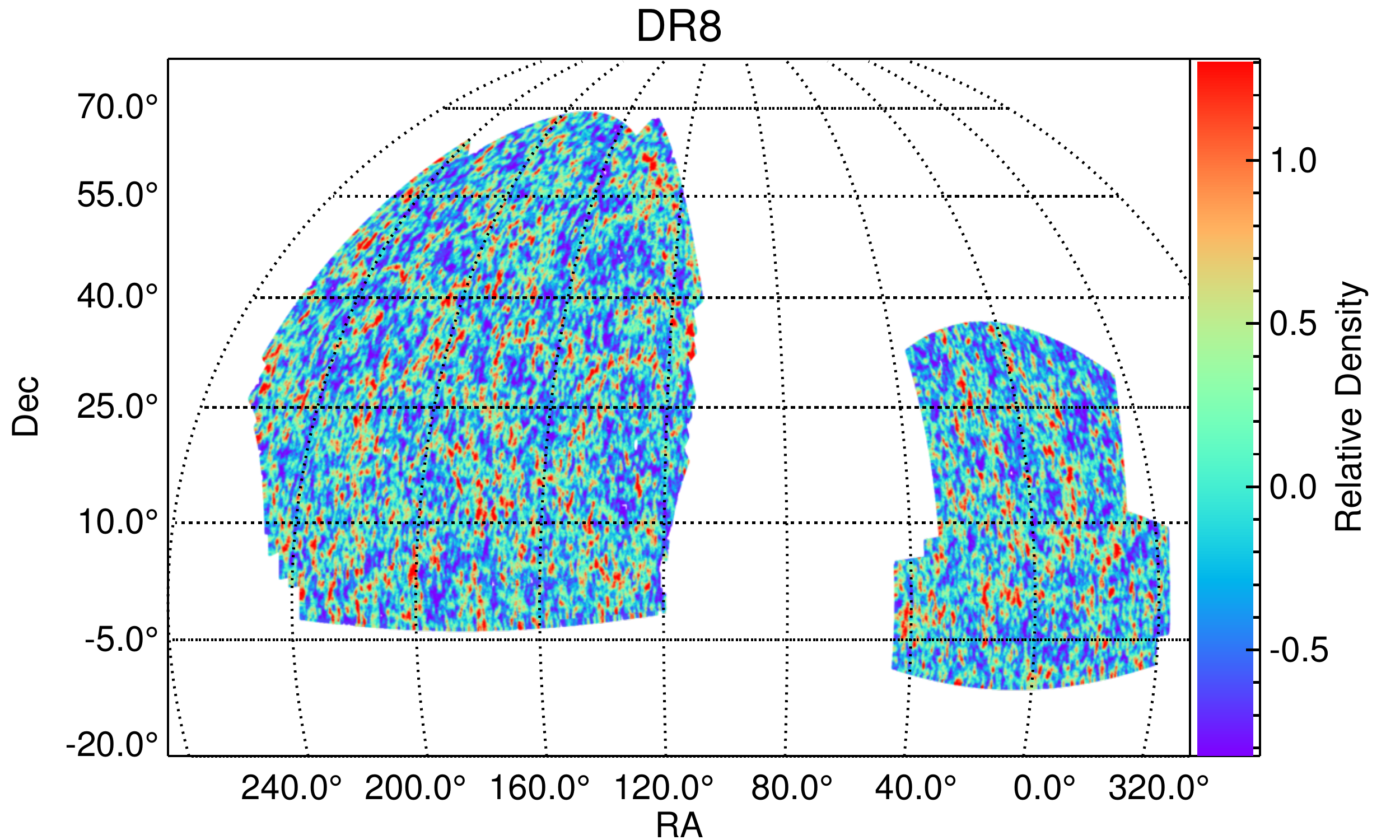}
  \caption{Angular cluster density contrast $\delta = (\rho -
    \bar{\rho})/\bar{\rho}$ for the SDSS DR8 \redmapper{} catalog in the
    redshift range $[0.1,0.3]$, averaged on a $30'$ scale.}
  \label{fig:dr8footprint}
\end{figure*}

\begin{figure*}
  \plotone{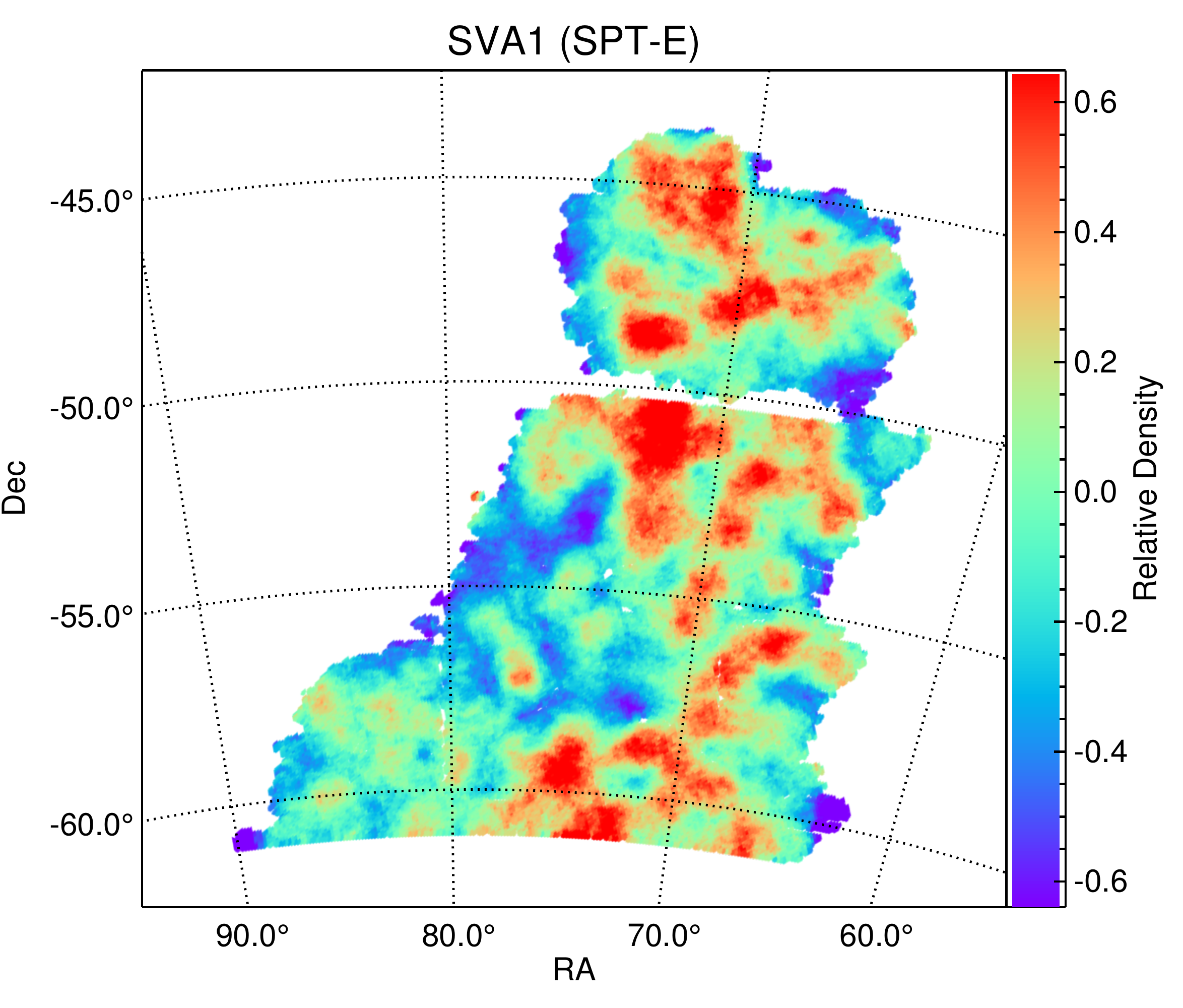}
  \caption{Angular cluster density contrast $\delta = (\rho -
    \bar{\rho})/\bar{\rho}$ for the DES SVA1 \redmapper{} catalog in the
    redshift range $[0.2,0.8]$, averaged on a $30'$ scale. }
  \label{fig:sva1footprint}
\end{figure*}

Figure~\ref{fig:dr8footprint} shows the angular density contrast of our
\redmapper{} sample for SDSS DR8 ($0.1<\zlambda<0.3$), and
Figure~\ref{fig:sva1footprint} shows the same for DES SVA1
($0.2<\zlambda<0.8$).  We restrict ourselves to $\zlambda<0.8$ because only the
deepest regions (and SN fields) have \redmapper{} selected clusters at
$\zlambda>0.8$.  Due to the relatively small density of clusters on the sky,
the density contrast is smoothed on a $30'$ scale to suppress noise.  Large
scale structure is readily apparent in the cluster density.  Previous DES work
has shown that the density field of \redmapper\ clusters is well correlated
with the underlying matter density field as determined from weak lensing
measurements \citep{changetal15b,vikrametal15}.

\subsection{Photo-$z$ Performance}

\subsubsection{SDSS DR8}

In Figure~\ref{fig:dr8photoz} we compare the photometric redshift $\zlambda$ to
the spectroscopic redshift of the CG (where available) for all clusters in DR8
with $\lambda>20$.  The top panel shows a density map of the
$\zspec$--$\zlambda$ relation, with $4\sigma$ outliers (such that
$|(\zspec-\zlambda)/\sigma_{\zlambda}| > 4$), which make up $1.1\%$ of the
population, marked as red points. The outlier clump at $\zlambda\sim0.4$ is
due to cluster miscentering rather than photometric redshift failures.
In RM1, we demonstrated that this clump of outliers is due to errors
in cluster centering rather than photometric redshift estimation.  Specifically,
these outliers represent clusters in which the photometrically assigned central
galaxy has a spectroscopic redshift that is inconsistent not only with the photometric
redshift of the cluster, but also the spectroscopic redshift of the remaining cluster 
members (see Figure 10 in RM1).  This failure mode is particularly pronounced near
filter transitions.  The bottom panel shows the bias (magenta dot-dashed line)
and scatter (cyan dot-dot-dashed line) about the 1--1 line (blue dashes).  The
performance is equivalent to that from RM1, with $\sigma_z/(1+z)<0.01$ over
most of the redshift range.

\subsubsection{DES SVA1}
\label{sec:sva1photozperf}

\begin{figure}
  \scalebox{1.2}{\plotone{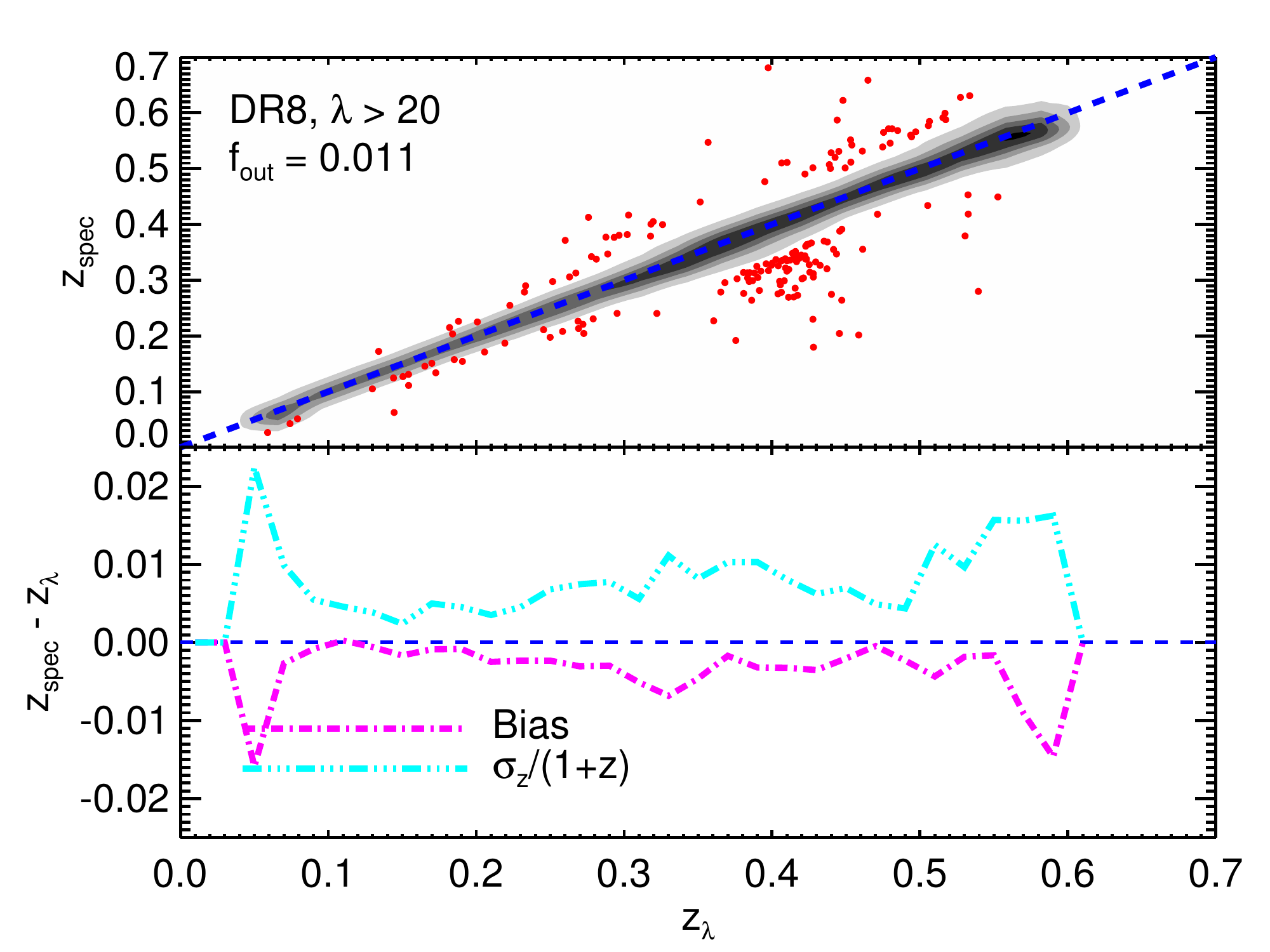}}
  \caption{\emph{Top:} Central galaxy spectroscopic redshift $\zspec$
    vs. cluster photometric redshift $\zlambda$ for SDSS DR8 clusters with
    $\lambda>20$.  Grey shaded regions show 1, 2, and $3\sigma$ density
    contours.  Red points, comprising $1.1\%$ of the total sample, show
    $>4\sigma$ outliers.  The outlier clump at $\zlambda\sim0.4$ is 
    not due to photometric redshift failures, but rather centering failures:
    these are primarily clusters with a correct photometric redshift, but whose
    photometrically assigned central galaxy is not in fact a cluster member.
    \emph{Bottom:} Bias in $\zspec - \zlambda$ (magenta) and
    $\zlambda$ scatter $\sigma_z/(1+z)$ (cyan) for clusters with central galaxy
    spectra.  Over most of the redshift range the bias is $<0.005$ and the
    scatter $\sigma_z/(1+z) < 0.01$.}
  \label{fig:dr8photoz}
\end{figure}

\begin{figure}
  \scalebox{1.2}{\plotone{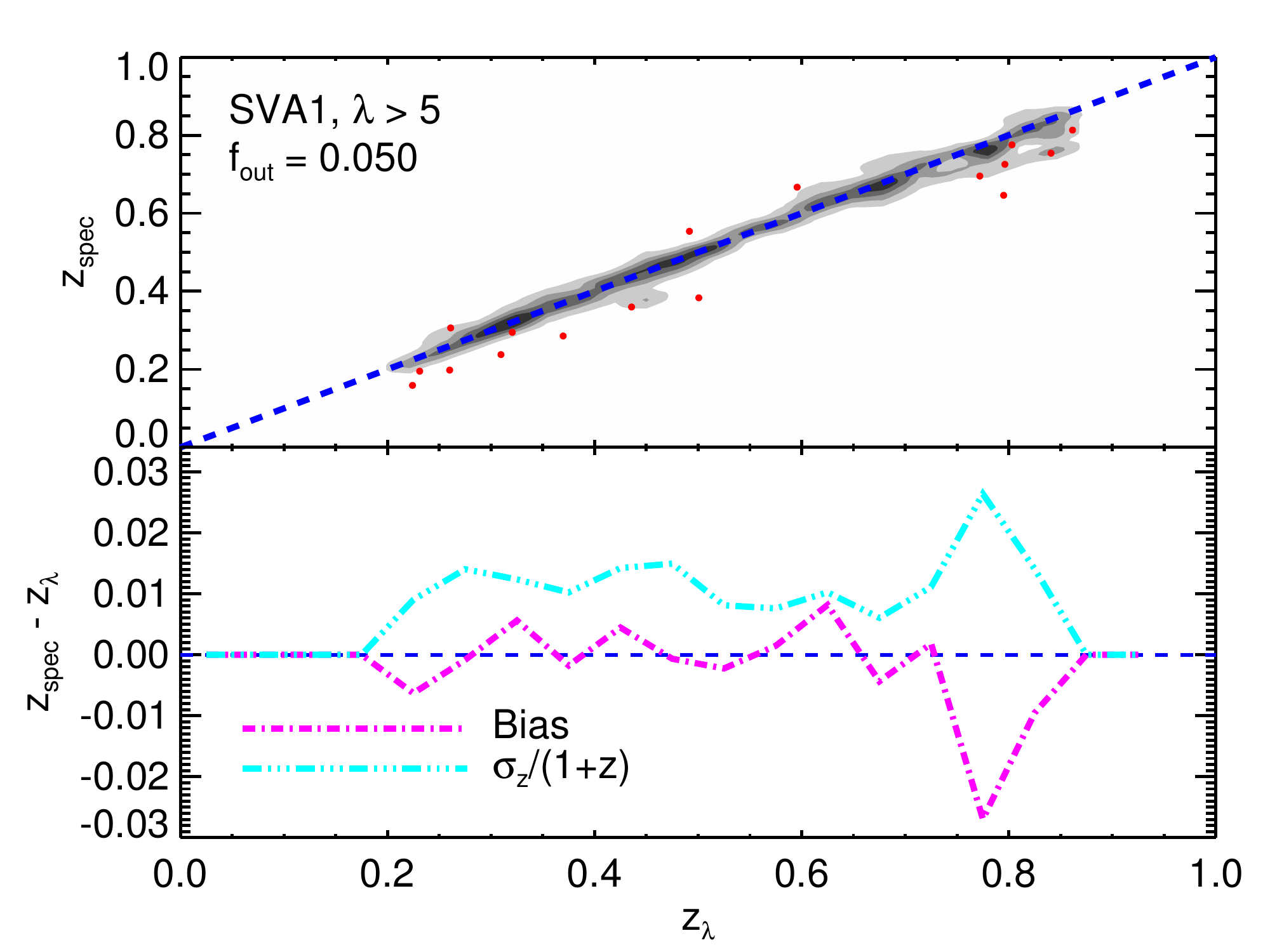}}
  \caption{Same as Figure~\ref{fig:dr8photoz}, for SVA1 clusters with
    $\lambda>5$.  The lower richness threshold was used for the plot because of
    the small number of cluster spectra for $\lambda>20$ clusters.  At
    $z\gtrsim0.7$ the scatter increases to $\sigma_z/(1+z)\sim0.02$ as our
    red-sequence model is noisy due to the relative lack of training
    spectra. As discussed in the text, the increased $\zlambda$ scatter over
    all redshifts (relative to DR8) is caused by relatively noisy \magauto{} colors.}
  \label{fig:sva1photoz}
\end{figure}

Figure~\ref{fig:sva1photoz} is the analogue to Figure~\ref{fig:dr8photoz} for
DES SVA1.  Because of the significantly smaller number of spectra, we show all
clusters with $\lambda>5$, despite the fact that this will increase the rate of
$4\sigma$ outliers due to miscentering.  Nevertheless, the performance is still
very good with only $5\%$ outliers.  All of these outliers have $\lambda<20$;
thus there are no $4\sigma$ outliers in the set of 52 clusters with spectra in
the fiducial $\lambda/S>20$ catalog.  The bias and scatter are all very good at
$z\lesssim0.7$, with an increase of $\sigma_z/(1+z)$ from $\sim0.01$ to
$\sim0.02$ at high redshift.  This increase is caused by both the variations in
survey depth, as well as noise in the high-$z$ red-sequence model that will be
reduced as we obtain more cluster spectra and increase our footprint in full
DES survey operations.  At low redshift, we note that the scatter in $\zlambda$
is larger in DES SVA1 than in SDSS DR8.  This is primarily caused by the
relatively noisy \magauto{} galaxy colors employed for our SVA1 catalog which
increase the red-sequence width and hence the noise in $\zlambda$.  

Because our analysis utilized all available spectroscopy for training
\redmapper, it is possible that our \photoz\ performance is artificially good
due to over-training.  To test for this, we have done a second full training of
the red-sequence model using only $50\%$ of the cluster spectra, and reserving
the second half for a validation test.  This is not ideal, as we then fall
below the required number of spectra for a good fit to the red-sequence model
(see Appendix B of RM1).  Nevertheless, the $\zlambda$ statistics of the validation
catalog are equivalent to those of the full fiducial run\footnote{Though the $\zlambda$
  statistics are the same, the richness estimations are not as stable, and thus
  our primary catalog utilizes all the spectra for training.}.

\subsection{Density of Clusters}
\label{sec:density}

\begin{figure}
  \scalebox{1.2}{\plotone{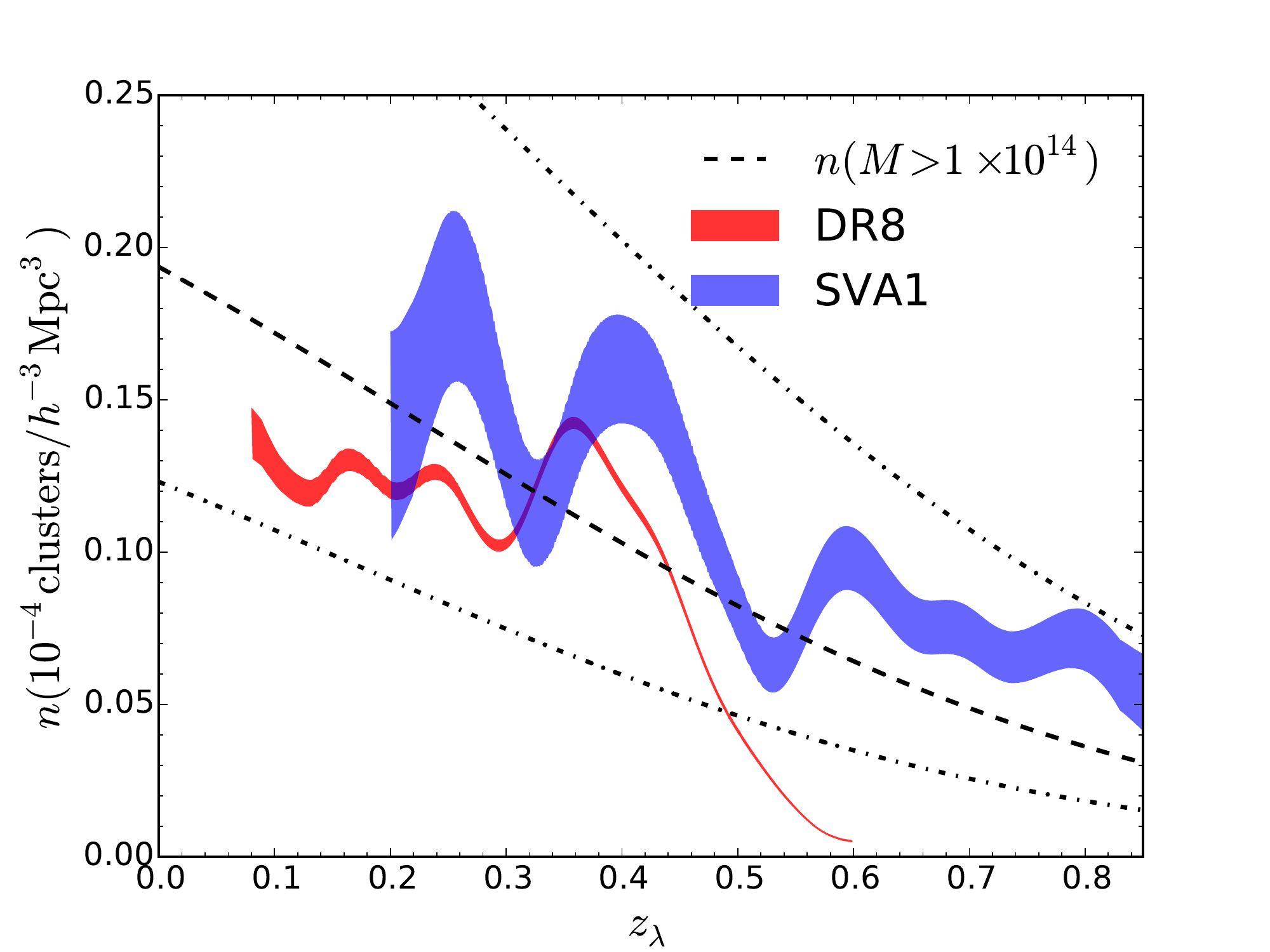}}
  \caption{Comoving density of clusters ($\lambda/S>20$) for DR8~(red curve)
    and SVA1~(blue curve), assuming our fiducial cosmology.  Width of the lines
    correspond to the assumption of Poisson errors (which are consistent with
    jackknife errors). The black dashed line shows the predicted abundance of
    halos with $M_{500c} > 1\times10^{14}\,h_{70}^{-1}\,\msun$, with the dash-dotted
    lines showing the same with a mass threshold of
    $0.7\times10^{14}\,h_{70}^{-1}\,\msun$ and
    $1.3\times10^{14}\,h_{70}^{-1}\,\msun$~\citep{tinkeretal08}.}
  \label{fig:comovingdensity}
\end{figure}

In Figure~\ref{fig:comovingdensity} we show the comoving density of
\redmapper{} clusters for DR8 (red) and SVA1 (blue).  Densities are computed
using our fiducial cosmology for clusters with $\lambda/S>20$ by summing
individual cluster $P(z)$ functions.  The width of the lines are smoothed over
a redshift range $\delta z=0.02$, and assume Poisson errors (which are
consistent with jackknife errors).
The black dashed line shows the predicted
abundance of halos with $M_{500c} > 1\times10^{14}\,h_{70}^{-1}\,\msun$, with the
dash-dotted lines showing the same with a mass threshold of
$0.7\times10^{14}\,h_{70}^{-1}\,\msun$ and
$1.3\times10^{14}\,h_{70}^{-1}\,\msun$~\citep{tinkeretal08}.  

We note that the \redmapper\ cluster is volume limited only out to $z\leq 0.33$.  Above this redshift, the 
cluster density as a function of redshift reflects two competing trends: an increasing \citet{eddington1913} bias in
the estimated cluster richness, which tends to increase the cluster density as a function of richness,
and an increasing detection threshold due to the shallow survey depth of the SDSS. 
For $z \approx 0.4$, the number of galaxies lost due to the shallow survey depth is relatively small,
and Eddington bias dominates, leading to an apparent increase in the cluster density.
As one moves towards even higher redshifts, the increasing detection threshold quickly dominates,
and the density of clusters falls as an increasing function of redshift.

The SVA1 density is
roughly consistent with DR8 at low redshift, although the volume probed is much
smaller; the peak at $z\sim0.6$ is caused by the same Eddington bias
effects as in DR8 at lower redshift.  The number density slowly declines with
redshift in SVA1, which is consistent with a constant mass threshold at
fixed richness.  However, we caution that the possibility of a varying mass
threshold (due to the build-up of the red-sequence, for example) as well as
Eddington bias and projection effects must both be taken into account to compute a proper cluster
abundance function  $n(z,M)$ for cosmological studies.

\section{Effects of Star/Galaxy Separation and Masking in SVA1}
\label{sec:sgmask}

As discussed in Section~\ref{sec:sva1data}, the fiducial SVA1 \redmapper{}
footprint was based on the area used for the \ngmix{} galaxy shape catalog, in
order to utilize the improved morphological star/galaxy separation in this
region.  In addition, we removed $4\%$ of the area with a relatively large
concentration of centroid shifts between bandpasses in individual objects.
However, these two choices come with some trade-offs.  While the improvement in
star/galaxy separation is clearly necessary in the selection of \redmagic{} red
galaxies (see Appendix A of RM15), it significantly reduced the footprint of
the SVA1 \redmapper{} catalog.  This is especially detrimental for the purposes
of comparing the \redmapper\ catalog against external X-ray cluster catalogs
(see Section~\ref{sec:xray}).  Similarly, while the bad region mask is clearly
beneficial for shape measurements, it creates a footprint with many holes,
which negatively impacts cluster centering.  In this section, we investigate
the impact of these choices on the richness and redshift recovery of
\redmapper{} clusters.  We also describe an expanded \redmapper{} catalog with
a larger footprint that can be used for multi-wavelength cross-correlation
measurements, increasing the number of clusters available in
Section~\ref{sec:xray} by $\sim50\%$.  

\subsection{Star/Galaxy Separation}
\label{sec:stargal}

The initial star/galaxy classifier in SVA1 data is the \modest{} classifier,
based on the {\tt SExtractor} {\tt SPREAD\_MODEL} quantity~\citep[][Section
  2.2]{changetal15, jarvisetal15} which compares the fit of a PSF model to that
of a PSF convolved with a small circular exponential model for morphological classification.  While \modest{}
works reasonably well at bright magnitudes, at $z\sim0.7$ the stellar locus (in
the DES optical bands $griz$) comes close to the galaxy red sequence.  For accurate
selection of individual red galaxies as in the \redmagic{} catalog, this
required our improved star/galaxy classification based on
\ngmix~\citep{redmagic15}, which reduced stellar contamination from $\gtrsim15\%$
at $z\sim0.7$ to less than $5\%$.

In order to estimate the impact of star/galaxy separation, we have rerun the
\redmapper{} cluster finder on a slightly expanded footprint using the \modest{}
star/galaxy classifier, while leaving everything else (including the
red-sequence calibration) the same.  We then match clusters from this 
catalog to our fiducial catalog.  The first thing we find is that a small
number of clusters ($\sim 1.4\%$) are now badly miscentered on bright, red,
misclassified stars (as determined from our improved star/galaxy separation
from \ngmix).  We also notice that the global background is slightly increased
at high redshift, thus slightly depressing the richness estimates.  The
richness bias is $\sim3\%$ at $z=0.8$, with the bias decreasing linearly with
redshift such that the cluster richnesses at $z=0.2$ are unbiased.  We
calibrate this bias with a simple linear model, and correct for it in our final
expanded catalog.  The associated systematic uncertainty in richness due to the
inefficient star/galaxy separation is $\sim 2\%$, smaller than the statistical
uncertainty on $\lambda$.  Thus, aside from mild miscentering problems, \redmapper{}
richness estimates are quite insensitive to stellar contamination in the galaxy
catalog, as expected.

\subsection{Masking}
\label{sec:masking}

In addition to the overall geometric mask, our fiducial footprint includes
masking for bright ($J<13$) 2MASS stars and $4\%$ of the area with a
larger-than-typical concentration of object centroid shifts.  However, we have
found that several good cluster centers are masked in these regions causing
significant offsets from the X-ray and SZ centers~(e.g., Section 2.3 of S15).

In order to estimate the impact of masking (in addition to star/galaxy
separation), we have rerun \redmapper{} on the expanded footprint using the
\modest{} classifier (as above), and including galaxies that had been rejected
by both the 2MASS mask and the ``$4\%$'' mask.  We then match clusters from
this expanded catalog to the fiducial catalog.  Aside from the clusters that
are now badly miscentered due to stellar contamination, two SPT
clusters~(SPT-CLJ0417$-$4748 and SPT-CL0456$-$5116; see S15) are now properly
centered as the central galaxies are no longer masked.

\begin{figure}
  \plotone{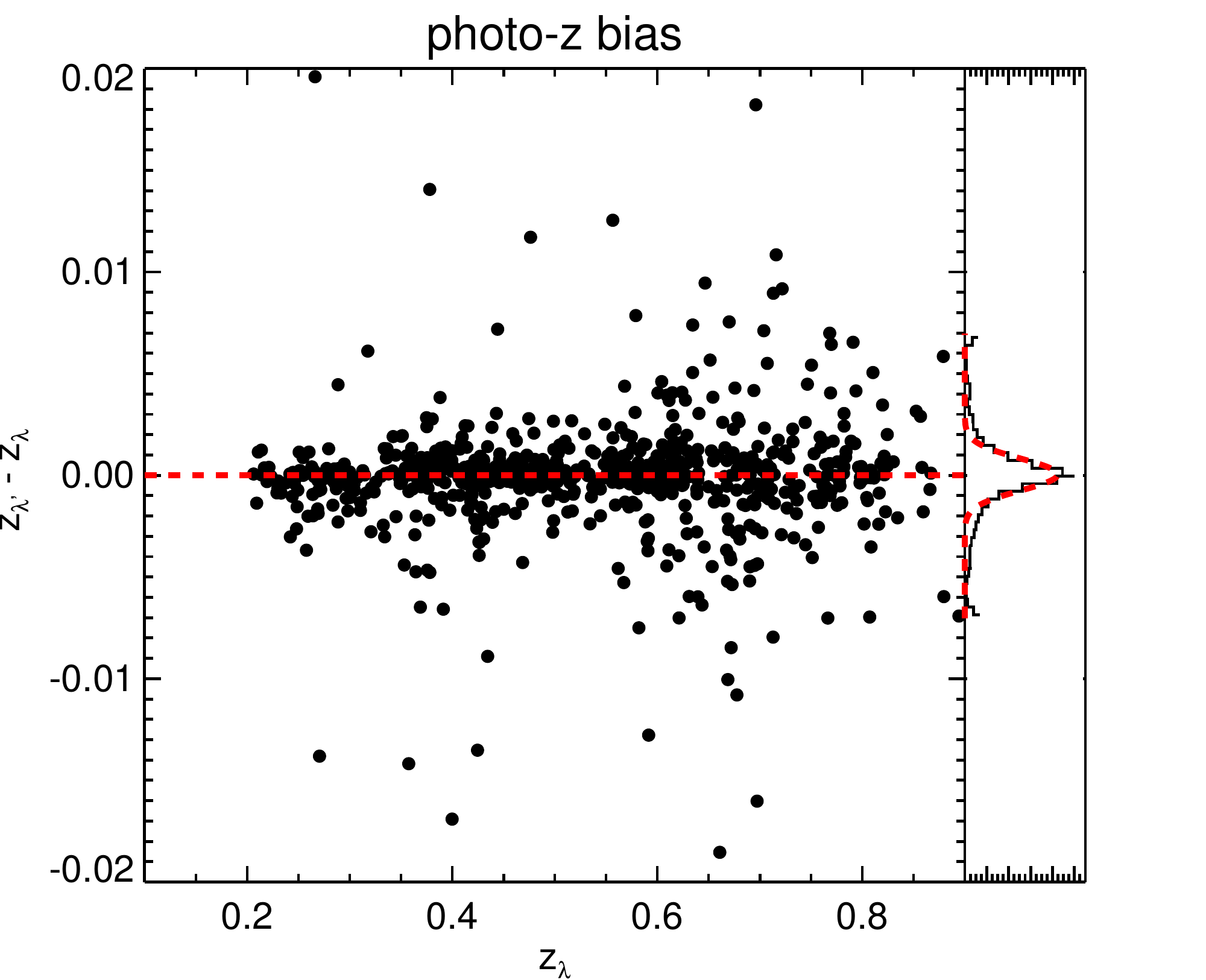}
  \caption{Plot of $\Delta\zlambda = z_{\lambda'} - \zlambda$ for the expanded
    ($z_{\lambda'}$) and fiducial ($\zlambda$) catalogs.  The cluster redshifts
    are very consistent, with few outliers at $\Delta\zlambda > 0.01$, which is
    already $<1\sigma$ on the redshift error.  The red dashed curve in the right panel
    is a Gaussian fit to the $\Delta\zlambda$ histogram, with mean
    $5\times10^{-5}$ and RMS $7\times10^{-4}$.}
  \label{fig:zlambda_expanded}
\end{figure}

Figure~\ref{fig:zlambda_expanded} shows the comparison in cluster redshift
$\zlambda$ between the expanded ($z_{\lambda'}$) and fiducial ($\zlambda$)
catalogs.  The cluster redshifts are very consistent, with a few outliers at
$\Delta\zlambda > 0.01$.  The red curve in the right panel shows a Gaussian fit to
the $\Delta\zlambda$ histogram, with mean $5\times10^{-5}$ and RMS
$7\times10^{-4}$.  Thus, the worse star/galaxy separation and less conservative
mask have no significant impact on the cluster redshift estimation.

\begin{figure}
  \plotone{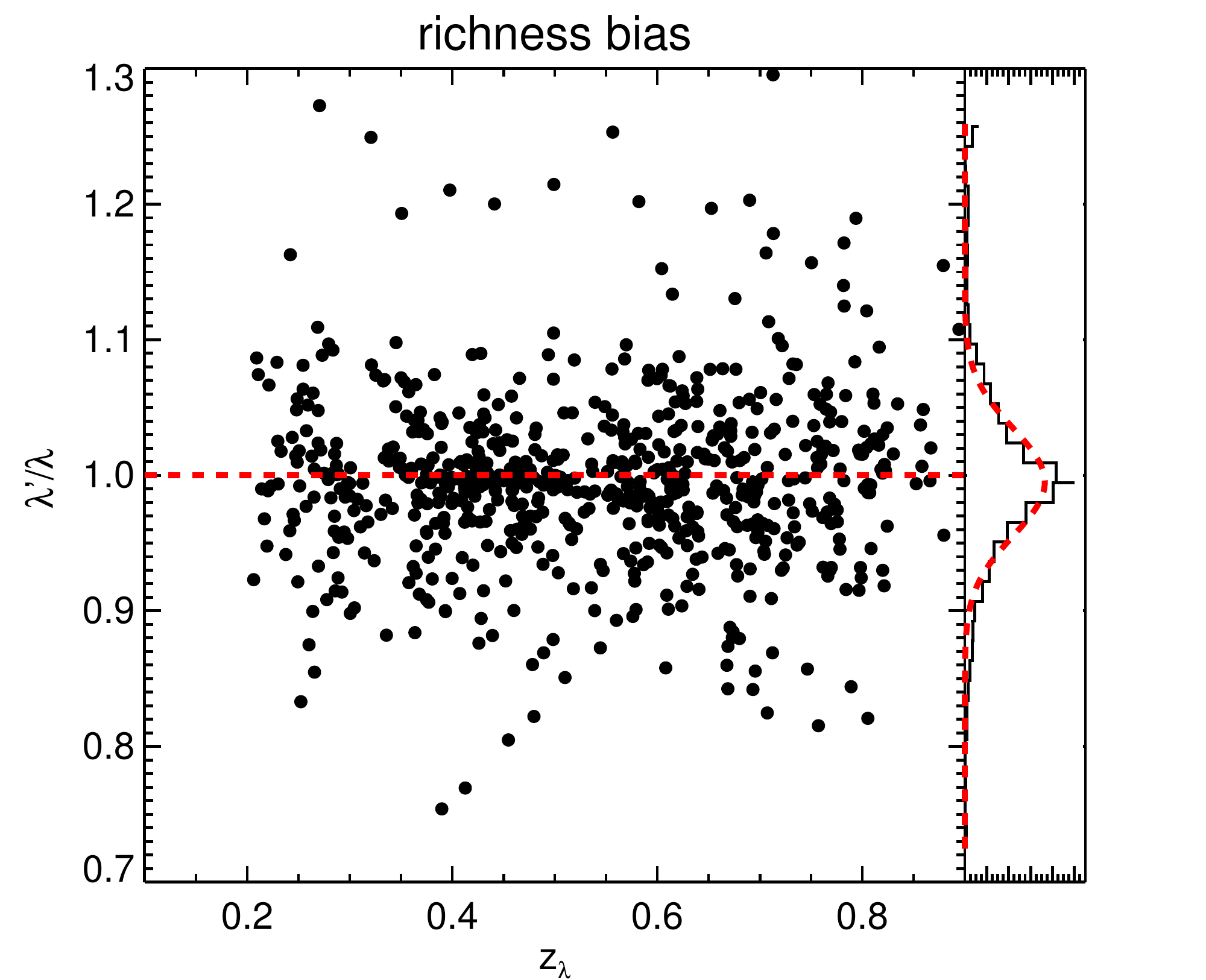}
  \caption{Plot of richness bias, $\lambda'/\lambda$, for the expanded ($\lambda'$) and
    fiducial ($\lambda$) catalogs.  All values of $\lambda'$ have been
    corrected for the star/galaxy separation bias model in
    Section~\ref{sec:stargal}.  The richness estimates are consistent, with a
    Gaussian fit (red dashed curve) showing $\lambda'/\lambda = 0.99\pm0.04$.}
  \label{fig:lambda_expanded}
\end{figure}

Figure~\ref{fig:lambda_expanded} shows the richness bias as the ratio of
$\lambda'$ (expanded catalog) to $\lambda$ (fiducial catalog) in the SPT-E
region.  All values of $\lambda'$ have been corrected for the star/galaxy
separation bias model in Section~\ref{sec:stargal}.  Again, the richness
estimates are consistent, with a Gaussian fit showing $\lambda'/\lambda =
0.99\pm0.04$.  We note that this $\sim 4\%$ richness scatter is fully
consistent with the expectations based on the richness extrapolations in the
fiducial catalog which made use of a more aggressive mask.  However, we also find that for
$\sim7\%$ of clusters $\lambda'/\lambda$ differs from unity by more than
$3\sigma$.  These apparent outliers are caused by clusters seen in projection. 
Changes in masking can change the way these projected clusters are 
deblended or merged by the \redmapper{} algorithm, leading to these outliers.  
This result suggests a lower limit of $\approx 7\%$ for the \redmapper\ projection rate,
and demonstrates the need for a full model of projection effects incorporated
into a cluster abundance function.

\begin{figure}
  \plotone{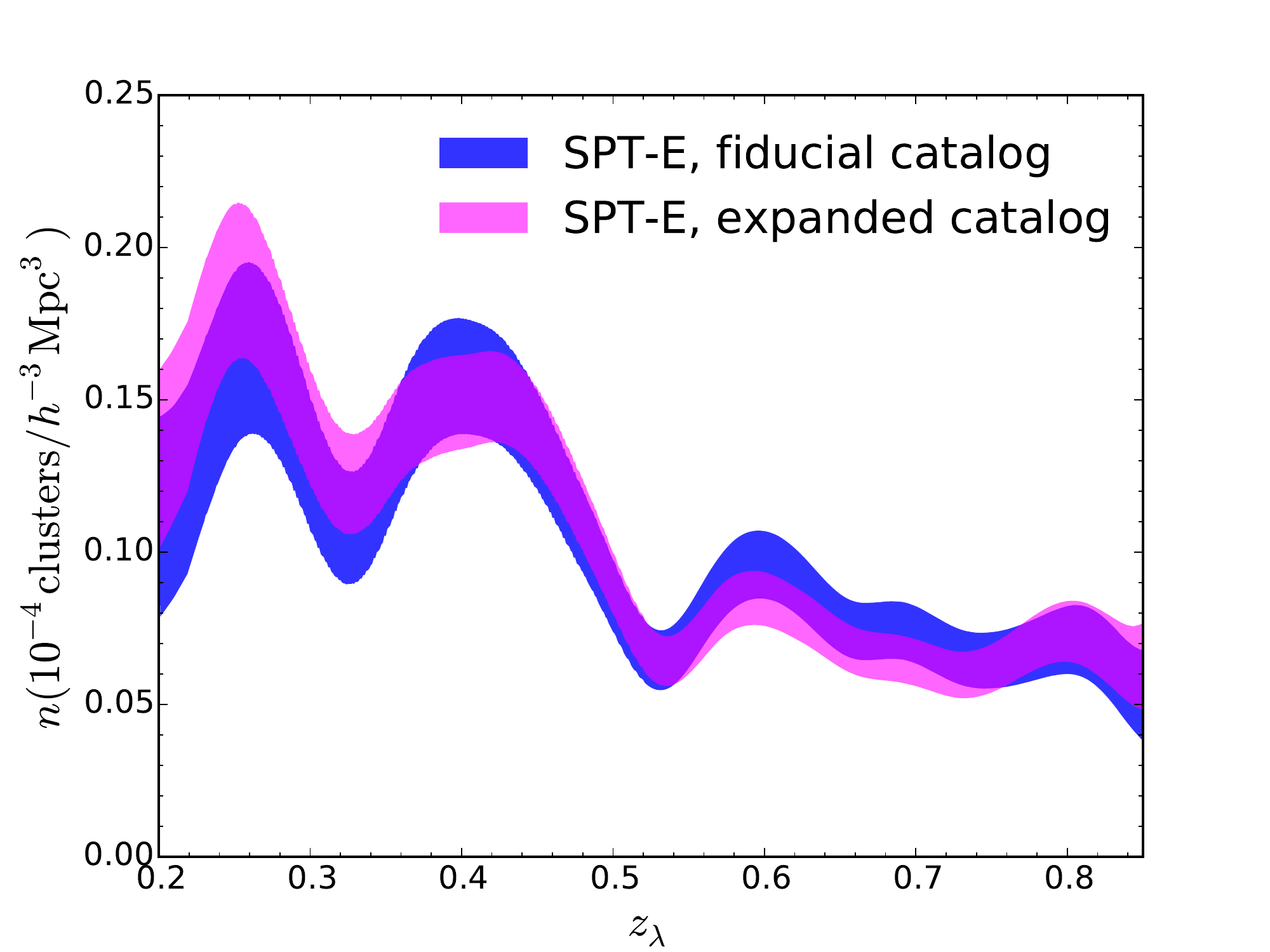}
  \caption{Number density of clusters for the expanded (magenta) and fiducial
    (blue) catalogs, limited to the SPT-E region.  The number density is
    consistent within $1\sigma$ at all redshifts in spite of the changes in
    star/galaxy separation and masking.}
  \label{fig:density_compare}
\end{figure}

In Figure~\ref{fig:density_compare} we show the comoving density of clusters in
the SPT-E region for our fiducial (blue) and expanded (magenta) catalogs.  The
number densities are clearly consistent at all redshifts.  Therefore, in future
versions of \redmapper{} on DES data our fiducial runs will be performed with a
less aggressive mask (with more area) as it has no impact in the richness
estimation, yet it does improve cluster centering in a small number of cases.
While improved star/galaxy separation is helpful for many purposes, it is
heartening to know that our richness estimates are not strongly biased by a
less-than-ideal separator.  For this version of the catalog, however, we
recommend that the fiducial catalog should be used for all purposes except
where the greater area can be made use of in cross-checks with X-ray catalogs,
as in Section~\ref{sec:xraysz}.

\section{The Correlation of \redmapper\ Cluster Richness with X-ray
and SZ Galaxy Cluster Properties}
\label{sec:xraysz}

\subsection{Correlation with the SPT SZ Cluster Catalog}

A detailed comparison of the DES SVA1 \redmapper{} and SPT SZ cluster catalogs
has been published in S15.  We briefly summarize their most
important results.  Using $129\,\mathrm{deg}^2$ of overlapping data, they find
25 clusters between $0.1<z<0.8$, including 3 new clusters that did not have
identified optical counterparts in \citet{bleemetal15}.  Every SZ cluster
within the \redmapper{} footprint and at $z<z_{\mathrm{max}}$ was detected in
the \redmapper{} catalog.  Due to the high mass
threshold of the \citet{bleemetal15} sample, these are all high-mass and
high-richness clusters, with a typical richness $\lambda\sim70$.  Using the
method of \citet{bocquetetal15}, they implement a full likelihood formalism to
constrain the $\lambda$--mass relation of SPT-selected clusters.  By inverting
the scaling relation from S15 using the methods of
\citet{evrardetal14}, they compute that the mass of a $\lambda\sim20$ cluster is
$M_{500c} \sim10^{14}\,h_{70}^{-1}\,\msun$, consistent with the density of
clusters from Section~\ref{sec:density}.   In addition, they find a
mass scatter at fixed richness, $\sigma_{\mathrm{ln}M|\lambda} =
0.18^{+0.08}_{-0.05}$, at a richness of $\lambda=70$.  Thus, they confirm that
the \redmapper{} richness $\lambda$ is a low-scatter mass proxy for DES data,
across a much broader range of redshift than was probed in~\citet{RM2}.
Furthermore, the parameters of the $\lambda$--mass relation are consistent with
what was derived from SDSS DR8 data using a rough abundance matching
argument~\citep{rykoffetal12}, thus giving further confirmation to the fact
that \redmapper{} is probing a similar cluster population in SDSS and DES data.

In S15, they further constrain the optical-SZE positional offsets.  The offset
distribution is characterized by a two-component Gaussian model.  The central
component describes ``well-centered'' clusters where the optical and SZ
positions are coincident (given the SZ positional uncertainty from the finite
beam size of SPT).  There is also a less populated tail of central galaxies
with large offsets.  For this work, we have modified the model such that the
central Gaussian component is a one-dimensional rather than 2D Gaussian, as we
have found this produces superior $\chi^2$ fits to the X-ray offsets in
Section~\ref{sec:xray}.  The positional offsets, $x$, are now modeled as:
\begin{equation}
  p(x) = \frac{\rho_0}{\sigma_0\sqrt{2\pi}}e^{-\frac{x^2}{2\sigma_0^2}} + \frac{(1-\rho_0)x}{\sigma_1^2}e^{-\frac{x^2}{2\sigma_1^2}} 
  \label{eqn:centeringmodel}
\end{equation}
where $x = r/R_{\lambda}$, $\rho_0$ is the fraction of the population with small
offsets with variance $\sigma_0^2$, and the population with large offsets is
characterized with variance $\sigma_1^2$.  We have refit the offset model of
SPT clusters from S15 $r/R_{\lambda}$
rather than $r/R_{500}$, in addition to using the \redmapper{} positions from
the expanded SVA1 catalog.  This change allows us to better compare to the X-ray
cluster samples described in Section~\ref{sec:xray}.  In all cases we
marginalize over the parameter $\sigma_0$ since it is not relevant to the
overall fraction and distribution of incorrect central galaxies.

The optical-SZ positional offset distribution has a central component with
$\rho_0 = 0.80^{+0.15}_{-0.37}$ and a large-offset population with
$\sigma_1=0.27^{+0.21}_{-0.08}R_{\lambda}$.  Given the matched clusters, the
mean of the centering probability of the central galaxies of the
clusters in the matched sample is $\langle\Pcen\rangle = 0.82$.  This is
consistent with the constraints from the optical-SZ matching, although the 21
clusters in the sample do not have a lot of constraining power.

\subsection{Correlation with X-ray Galaxy Clusters}
\label{sec:xray}

In this section, we make use of the overlap of the \redmapper{} SVA1 expanded
catalog with X-ray observations from \emph{Chandra} and \emph{XMM} to measure
the $T_X$--$\lambda$ relation, as well as further constrain the centering
properties of the catalog.  More extensive comparisons to X-ray observations,
including a full analysis of the \redmapper{} DR8 catalog, will be presented in
Hollowood et al. (in prep) and Bermeo Hernandez et al. (in prep).

\subsubsection{Chandra Analysis}
\label{sec:chandra}

The \emph{Chandra} analysis was performed using a custom pipeline (see
Hollowood et al., in preparation). A brief overview is given here.  The
pipeline is based on a series of CIAO (version 4.7)\citep{CIAO} and HEASOFT
(version 6.17) tools; all spectral fitting was performed using
\emph{XSPEC}~\citep[version 12.9.0,][]{XSPEC}.

The \emph{Chandra} pipeline was used to extract temperatures and luminosities
from a list of clusters that were both in the \redmapper{} catalog
($\lambda>20$) and in at least one \emph{Chandra} archival observation. 
The pipeline took a list of cluster positions, redshifts, and
richnesses from the \redmapper{} SVA1 expanded catalog, and queried the
\emph{Chandra} archive for observations of these positions using the
\emph{find\_chrandra\_obsid} CIAO tool. The pipeline then downloaded each
observation which contained a \redmapper{} cluster, and re-reduced it using the
\emph{chandra\_repro} CIAO tool.

Each observation was then cleaned using a standard X-ray analysis: the
energy was cut to 0.3--7.9 keV, flares were removed using the \emph{deflare}
CIAO tool with the \emph{lc\_clean} algorithm, and point sources were removed
using the \emph{wavdetect} CIAO tool. A $500\,\mathrm{kpc}$ radius was then
calculated around the \redmapper{} center, using the \redmapper{} redshift
$\zlambda$ and assuming a cosmology of $\Omega_m=0.3$, $H_0=0.7$. This region
was then iteratively recentered to the local X-ray centroid. At this point, the
signal-to-noise ratio in this region was measured, and if it was less than 3.0,
analysis stopped. Otherwise, a spectrum was extracted from this region.

A temperature was then fit to this spectrum using a {\tt WABS}$\times${\tt
  MEKAL} model~\citep{meweetal85}, fixing the Hydrogen column
density to the \citet{dickeyandlockman90} from the \emph{nH} HEASOFT tool, and
the metal abundance to 0.3 times Solar. An
$r_{2500}$ radius was derived from this temperature via the empirical relation
found in \citet{arnaudetal05}. The derived $r_{2500}$ radius was then used to
create an iteratively-centered $r_{2500}$ region, which was then used to
produce a new $r_{2500}$ temperature and radius. The temperature and radii were
then iterated until they converged within one sigma. Unabsorbed soft-band
(0.5--2.0 keV) and bolometric (0.001--100 keV) luminosities were then
calculated for the data. 

In the \redmapper{} expanded SVA1 sample, 61 clusters fell within a
\emph{Chandra} archival region, 38 of which had a sufficient signal-to-noise to
be analyzed. Of these 38 clusters, 15 had sufficient statistics to fit an
$r_{2500}$ temperature.  Finally, we reject one cluster from the comparison
where the X-ray centroid is in a region of the \redmapper{} footprint with
$\fmask > 0.2$.  The cluster positions and temperatures used in this work are
described in Table~\ref{tab:chandraclusters}.

\subsubsection{XCS Analysis}
\label{sec:xcs}

The \emph{XMM-Newton} (\emph{XMM}) analysis was performed using an adaption of
the pipeline developed for the \emph{XMM} Cluster
Survey~\citep[XCS;][]{mehrtensetal12}.  XCS uses all available data in the
\emph{XMM} public archive to search for galaxy clusters that were detected
serendipitously in \emph{XMM} images.  X-ray sources are detected in \emph{XMM}
images using an algorithm based on wavelet transforms~\citep[see][for details,
LD11 hereafter]{ld11}.  Sources are then compared to a model of the instrument
point spread function to determine if they are extended. Extended sources are
flagged as cluster candidates because most extended X-ray sources are clusters
(the remainder being low-redshift galaxies or supernova remnants).

We have matched all XCS cluster candidates within $1.5\,h^{-1}\,\mathrm{Mpc}$
of a \redmapper{} SVA1 cluster with $\lambda>5$ (assuming the candidate lies at
the \redmapper{} determined redshift), although we note that all the verified
matches were within $0.4\,h^{-1}\,\mathrm{Mpc}$.  We note that for this match,
the default XCS defined X-ray center was used (see LD11 for more information
about XCS centroiding). If multiple matches are made, only the closest match is
retained. The initial matched sample contains 66 objects that passed XCS
quality standards. An average X-ray temperature estimate for each cluster was then
calculated for these objects using a method very similar to that described in
LD11. The XCS $T_{X}$ pipeline uses XSPEC~\citep{XSPEC} to fit a {\tt
  WABS}$\times${\tt MEKAL} model~\citep{meweetal85}, fixing the Hydrogen column
density to the \citet{dickeyandlockman90} value and the metal abundance to 0.3
times the Solar value. For the study presented herein, differences compared to
the LD11 version of the pipeline include the use of updated \emph{XMM}
calibration and XSPEC (12.8.1g) versions, and the extraction of $T_{X}$ values
within $r_{2500}$ regions.  We compute $r_{2500}$ using the same method as
Section~\ref{sec:chandra}.  Of the 66 matches between XCS cluster candidates
and \redmapper{} SVA1, we obtain $T_{X,2500}$ values for 31, with the
remainding clusters detected with insufficient signal to noise.  We have
checked the SVA1 images of each of these 31, with and without \emph{XMM} flux
contours overlaid.  After doing so, we discarded 6 objects because the XCS to
\redmapper{} match was clearly serendipitous.  Finally, we select only those
clusters with $\lambda>20$, and we reject 4 clusters from the comparison where
the X-ray centroid is in a region of the \redmapper{} footprint with $\fmask >
0.2$.  Our final sample of XCS clusters with positions (29) and the subset with
$T_X$ estimates (14) used in this work are described in
Table~\ref{tab:xcsclusters}.

\subsubsection{The $T_X$--$\lambda$ Relation}
\label{sec:txlambda}
  
For this study we wished to determine the \redmapper{} $T_{X,2500}$--$\lambda$
relation using clusters with either \emph{XMM} or \emph{Chandra} observations.
However, it is well known that X-ray cluster temperatures derived from
\emph{XMM} are systematically offset from \emph{Chandra}
observations~\citep[e.g.,][]{schellenbergeretal15}.  Therefore, we have
determined a correction factor to make the \emph{Chandra} and \emph{XMM}
temperatures consistent.  For this, we required access to more \redmapper{}
clusters with X-ray observations than are available in DES SVA1.  We rely on
recent compilations of $T_X$ measurements of SDSS \redmapper{} clusters using
\emph{Chandra}~(Hollowood et al., in prep) and \emph{XMM}~(Bermeo Hernandez et al., in
prep).  There are 41 DR8 \redmapper{} clusters in common between these samples,
allowing us to fit a correction factor of the form
\be
\log_{10}\left(\frac{T_{X}^{\mathrm{Chandra}}}{1\,{\rm
    keV}}\right)=1.0133\log_{10}\left(\frac{T_{X}^{\mathrm{XMM}}}{1\,\rm{keV}}\right)+0.1008
\label{eqn:txcorr}
\ee
using BCES orthogonal fitting~\citep{akritasandbershady96}.  We note that this
relation is consistent with that found by \citet{schellenbergeretal15}.  Of the
14 \redmapper{} SVA1 clusters with $T_{X,2500}^{\mathrm{XMM}}$ values, 4 are in common
with the \emph{Chandra} sample.  Of these 4, we have used the $T_{X,2500}$
value with the lowest uncertainty (3 from \emph{XMM} and 1 from
\emph{Chandra}).

\begin{figure}
  \scalebox{1.2}{\plotone{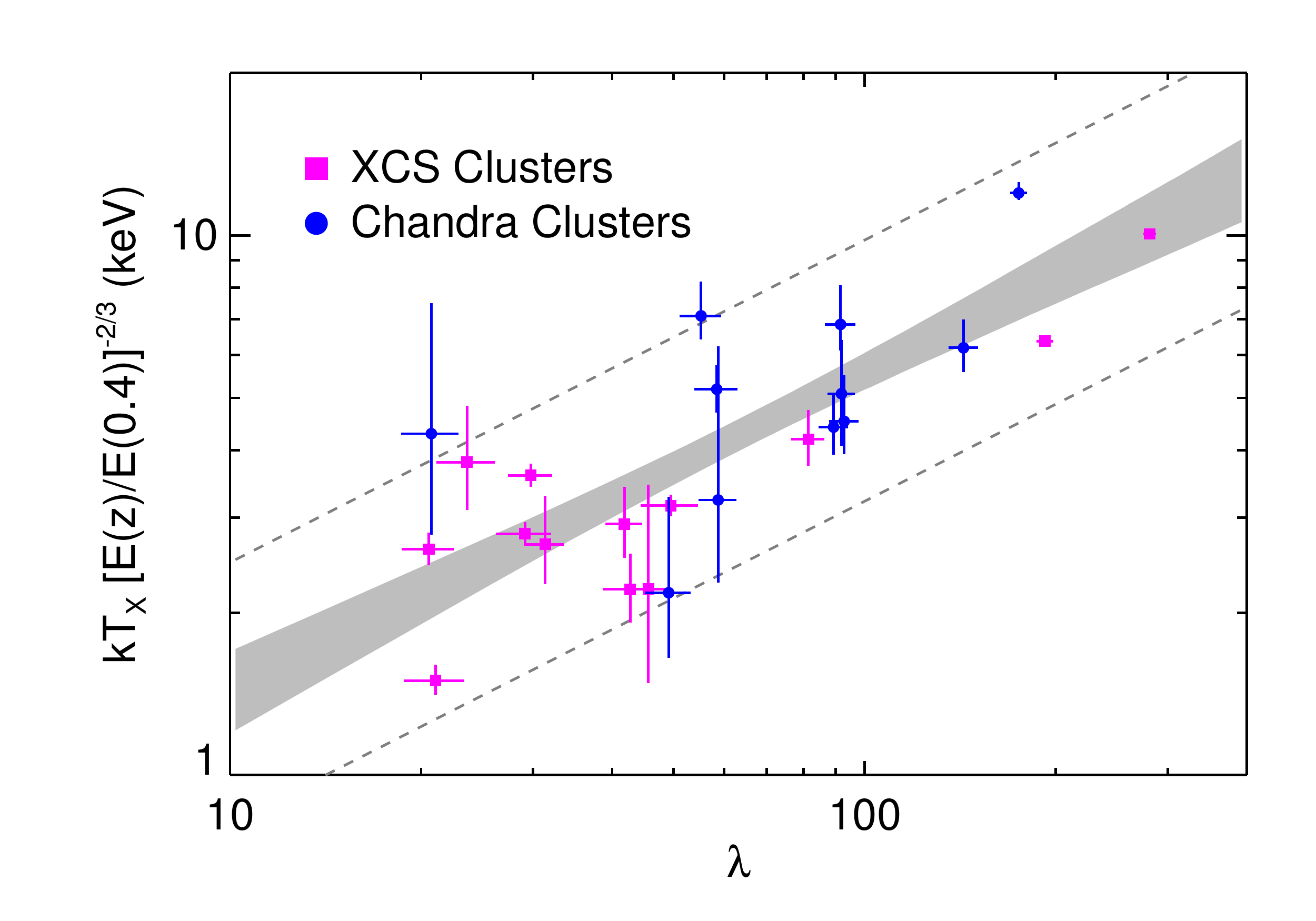}}
  \caption{$T_X$--$\lambda$ scaling relation derived from XCS (magneta squares)
    and \emph{Chandra} (blue circles) clusters.  All \emph{Chandra}
    temperatures have been corrected according to Eqn.~\ref{eqn:txcorr}.  The
    gray band shows the best fit ($\pm1\sigma$) scaling relation, and the
    dashed gray lines show $2\sigma_{\mathrm{int}}$ intrinsic scatter
    constraints.}
  \label{fig:txlambda}
\end{figure}

Figure~\ref{fig:txlambda} shows the $T_X$--$\lambda$ scaling relation derived
from XCS and \emph{Chandra} clusters.  All \emph{Chandra} temperatures have
been corrected according to Eqn.~\ref{eqn:txcorr}.  We use an MCMC to fit the full
cluster sample to a power-law model:
\be
\ln(T_X) = \alpha + \beta \ln (\lambda/50) + \gamma \ln [E(z)/E(0.4)],
\label{eqn:txlambda}
\ee
with intrinsic scatter $\sigma_{\ln T|\lambda}$.  Given the limited number of
clusters in our sample, we fix the redshift evolution parameter $\gamma=-2/3$,
assuming self-similar evolution.  We find that $\alpha =
1.31\pm.07$, $\beta=0.60\pm0.09$, and $\sigma = 0.28^{+0.07}_{-0.05}$.  
The best-fit scaling relation (including $1\sigma$ error) is shown with the gray bar in
Figure~\ref{fig:txlambda}, and the dashed lines show the $\pm2\sigma_{\ln
  T|\lambda}$ constraints.  We note that the slope is consistent with
$\beta=2/3$, which is what we expect if clusters are self-similar and $\lambda
\propto M$, as in S15.
 
S15 used SZ-selected clusters to place a constraint on the scatter in mass at
fixed richness $\sigma_{\ln M|\lambda} = 0.18^{+0.08}_{-0.05}$.  To compare
against S15, we transform our constraints on the scatter in $T_X$ at fixed mass
to constraints on scatter in mass at fixed richness by assuming a self-similar
slope $T_X \propto M^{2/3}$.  We also require an estimate for the scatter in
mass at fixed X-ray temperature, when $T_X$ is not core-excised.  We rely on
the results by \citet{lieuetal15}, who find an intrinsic scatter in weak
lensing mass at fixed temperature of $\sigma_{\ln M_{WL}|T} = 0.41$.  Our
choice is motivated by the fact this study, like ours, measure cluster
temperatures with no core-excision.  Adopting a 25\% intrinsic scatter in weak
lensing mass at fixed mass, we arrive at an intrinsic scatter in mass at fixed
temperature $\sigma_{\ln M|T}=0.32$.  Finally, given that the X-ray cluster
sample extends to low richness systems, which are expected to have a larger
scatter, we adopt a richness dependent scatter as a function of mass:
\be
\mbox{Var}(\ln \lambda|M) = \langle \lambda|M\rangle^{-1} + \sigma_{\ln
  \lambda|M}^2.
\ee
Following \citet{evrardetal14}, we arrive at $\sigma_{\ln
  M|\lambda} = 0.3 \pm 0.15$.  This result is higher than but consistent with
that of S15.  The large error bars reflect in part the large intrinsic scatter
in the mass--$T_X$ relation for non-core-excised temperatures.

\subsubsection{Positional Offset Distribution}
\label{sec:posoff}

Using each of the SPT, \emph{Chandra}, and XCS \redmapper-matched samples, we
have fit the positional offset model of Eqn.~\ref{eqn:centeringmodel}.  For the
SZ sample, the error on the position was given by Eqn. 11 from
S15; for the X-ray samples, we used a fixed error of $10''$.
However, we note that this does not fully account for systematic errors in
X-ray centroids, especially for clusters with complex morphologies.  For the
X-ray samples, we do not require the cluster be bright enough to get a
temperature constraint in order for it to have a well-detected center.

The offset model results are summarized in Table~\ref{tab:offsets}, with errors
quoted as $68\%$ confidence intervals as derived from an Markov Chain Monte
Carlo fit to the data, similar to Section 4.3 of S15.  The
constraints on $\rho_0$ and $\sigma_1$ (the large-offset ``miscentered''
component) are all consistent within errors for all three samples. 

To better constrain the overall centering of the \redmapper{} SVA1 expanded
cluster sample, we have also performed a joint likelihood fit to all three
cluster samples.  In the cases where we have multiple observations of the
same cluster, we first take the XCS position, followed by the \emph{Chandra}
position, followed by the SPT position.  As our goal is to better constrain the
well-centered fraction $\rho_0$ as well as the miscentering kernel $\sigma_1$,
our joint likelihood constrains these two parameters for the full sample.
However, to allow for differences in centering precision, we use a separate
value of $\sigma_0$ for each individual sample.  In all, we have 5 parameters,
but we treat the set of $\{\sigma_0\}$ as nuisance parameters in our figures below.

\begin{figure}
  \scalebox{1.2}{\plotone{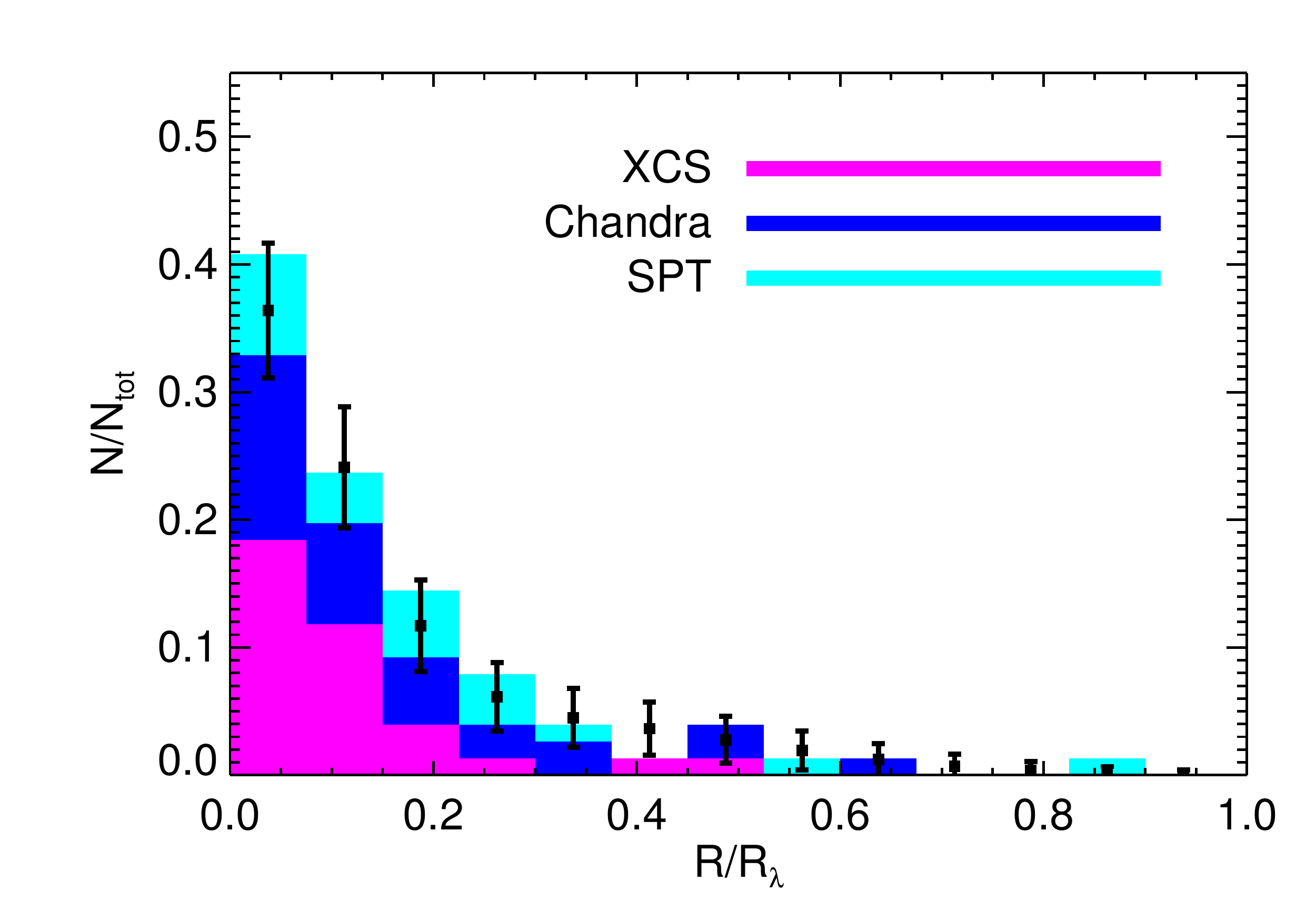}}
  \caption{Histogram of positional offsets for the combined cluster sample as a function
    of $R/R_{\lambda}$.  XCS clusters are shown in magenta, \emph{Chandra} clusters
    in blue, and SPT clusters in cyan.  The best-fit offset model, binned
    according to the data, is shown with black points. For reference, the average value of
    $\langle R_{\lambda} \rangle = 0.85\,h^{-1}\,\mathrm{Mpc}$, and the
    largest cluster offset is $0.75\,h^{-1}\,\mathrm{Mpc}$.}
  \label{fig:combhist}
\end{figure}

\begin{figure}
  \scalebox{1.2}{\plotone{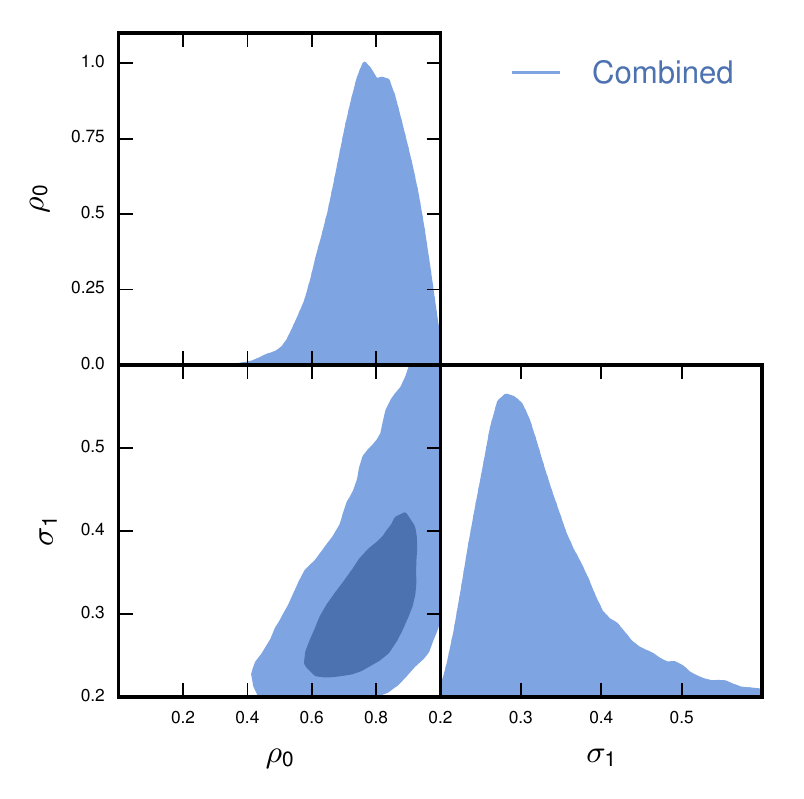}}
  \caption{Posterior distribution for the $1\sigma$ and $2\sigma$ levels of two
    of the parameters ($\rho0, \sigma_1$) from the positional offset model in
    Eqn.~\ref{eqn:centeringmodel}, for the combined XCS, \emph{Chandra}, and
    SPT data.  Best fit parameters are shown in Table~\ref{tab:offsets}.  The
    predicted well-centered fraction determined from \redmapper{} centering
    probabilities is $\langle\Pcen\rangle = 0.82$, consistent with the best-fit value.}
  \label{fig:comboffset}
\end{figure}

The histogram of offsets for the combined sample is shown in
Figure~\ref{fig:combhist}.  The results of our joint fit are shown in
Figure~\ref{fig:comboffset}, and described in Table~\ref{tab:offsets}.  The
best-fit model has been binned to match the data, and overplotted with black
points in Figure~\ref{fig:combhist}. Our final constraint on the fraction of
clusters that are correctly centered is $\rho_0=0.78^{+0.11}_{-0.11}$, compared
to the \redmapper{} predicted fraction of $0.82$, in very good agreement.  By
comparison, \redmapper{} clusters in SDSS are correctly centered $\approx 86\%$
of the time~\citep[see][]{RM2}.

\begin{deluxetable}{llllll}
\tablewidth{0pt}
\tablecaption{\redmapper Central Offset Fits}
\tablehead{
 \colhead{Sample} &
 \colhead{No.} &
 \colhead{$\langle\Pcen\rangle$} &
 \colhead{$\rho_0$} &
 \colhead{$\sigma_1 (R/R_{\lambda})$} & 
 \colhead{$\chi^2/\mathrm{DOF}$}
}
\startdata
SPT & 21 & $0.83$ & $0.80^{+0.15}_{-0.37}$ & $0.27^{+0.21}_{-0.08}$ & $6.0/10$\\
\emph{Chandra} & 35 & $0.80$ & $0.68_{-0.18}^{+0.22}$ & $0.27_{-0.05}^{+0.12}$
& $4.7/10$\\
XCS & 29 & $0.82$ & $0.85^{+0.07}_{-.11}$ & $0.22^{+0.08}_{-0.04}$ & $9.1/10$\\
Combined & 74  & $0.81$ & $0.78^{+0.11}_{-0.11}$ & $0.31^{+0.09}_{-0.05}$ & $9.9/10$\\
\enddata
\label{tab:offsets}
\end{deluxetable}

\section{Summary}
\label{sec:summary}

We present the DES SVA1 \redmapper{} cluster catalog, and an updated version of
the SDSS \redmapper\ cluster catalog.  Relative to the last \redmapper\ public
release (v5.10, see RM4), this new version (v6.3) includes a variety of improvements,
specifically:
\begin{enumerate}
\item{The algorithm now makes use of the depth maps generated as per \citet{rykoffdepth15}
  to properly account for small scale structure in the survey properties.}
\item{The synthetic curves for passive evolutions used by \redmapper{} are now
  internally generated using the BC03 model.}
\item{The selection of initial candidate red galaxies has been improved,
  allowing \redmapper{} to internally self calibrate with sparser spectroscopic
  data.}
\item{The catalog has a position-dependent redshift reach determined from the
  underlying survey inhomogeneity.}
\item{We have updated our generation of random points to properly account for
  the above changes, particularly the position-dependent redshift reach of the
  cluster catalog.}
\end{enumerate}

As with previous releases, the photometric redshift performance of the SDSS
catalogs is superb, being nearly unbiased and with photometric redshift scatter
$\sigma_z/(1+z) \leq 0.01$, except for the most distant clusters.  Photometric
redshift performance in DES SV is also excellent, with a scatter
$\sigma_z/(1+z)\approx 0.01$, only now the redshift range of the cluster
catalog extends to $z=0.9$.  The cluster richness has been shown to be tightly
correlated with cluster mass ($\approx 20\%$ scatter) by S15.
We have further validated this tight scatter using X-ray scaling relations.
These analysis, as well as the comoving density of galaxy clusters in DES SVA1,
suggests that the DES SVA1 detection threshold corresponds to a limiting mass
$M_{500c} \approx 10^{-14}\ h^{-1}\,\msun$ for our high-quality $\lambda>20$
cluster sample.

Finally, we have investigated the miscentering distribution of the DES SV
clusters.  The current data place only modest constraints on the miscentering
distribution, and we find that the fraction of clusters that are correctly
centered is $\approx 0.78\pm0.11$, fully consistent with our expectations from
the \redmapper{} centering probabilities, $\Pcen$.

Our results demonstrate that the DES imaging data is sufficiently robust and of
sufficient quality to pursue photometric cluster finding up to redshift
$z=0.9$, with well controlled selection functions, richness measurements, and
excellent photometric redshift performance, setting the stage for upcoming
analyses and cluster abundance constraints with the largest cluster samples
available to date.

\acknowledgments

This paper has gone through internal review by the DES collaboration.  We are
grateful for the extraordinary contributions of our CTIO colleagues and the
DECam Construction, Commissioning and Science Verification teams in achieving
the excellent instrument and telescope conditions that have made this work
possible.  The success of this project also relies critically on the expertise
and dedication of the DES Data Management group.

This work was supported in part by the U.S. Department of Energy contract to
SLAC no. DE-AC02-76SF00515, as well as DOE grants DE-SC0007093 (DH) and
DE-SC0013541 (DH and TJ).

Funding for the DES Projects has been provided by the U.S. Department of Energy, the U.S. National Science Foundation, the Ministry of Science and Education of Spain, 
the Science and Technology Facilities Council of the United Kingdom, the Higher Education Funding Council for England, the National Center for Supercomputing 
Applications at the University of Illinois at Urbana-Champaign, the Kavli Institute of Cosmological Physics at the University of Chicago, 
the Center for Cosmology and Astro-Particle Physics at the Ohio State University,
the Mitchell Institute for Fundamental Physics and Astronomy at Texas A\&M University, Financiadora de Estudos e Projetos, 
Funda{\c c}{\~a}o Carlos Chagas Filho de Amparo {\`a} Pesquisa do Estado do Rio de Janeiro, Conselho Nacional de Desenvolvimento Cient{\'i}fico e Tecnol{\'o}gico and 
the Minist{\'e}rio da Ci{\^e}ncia, Tecnologia e Inova{\c c}{\~a}o, the Deutsche Forschungsgemeinschaft and the Collaborating Institutions in the Dark Energy Survey. 

The Collaborating Institutions are Argonne National Laboratory, the University of California at Santa Cruz, the University of Cambridge, Centro de Investigaciones Energ{\'e}ticas, 
Medioambientales y Tecnol{\'o}gicas-Madrid, the University of Chicago, University College London, the DES-Brazil Consortium, the University of Edinburgh, 
the Eidgen{\"o}ssische Technische Hochschule (ETH) Z{\"u}rich, 
Fermi National Accelerator Laboratory, the University of Illinois at Urbana-Champaign, the Institut de Ci{\`e}ncies de l'Espai (IEEC/CSIC), 
the Institut de F{\'i}sica d'Altes Energies, Lawrence Berkeley National Laboratory, the Ludwig-Maximilians Universit{\"a}t M{\"u}nchen and the associated Excellence Cluster Universe, 
the University of Michigan, the National Optical Astronomy Observatory, the University of Nottingham, The Ohio State University, the University of Pennsylvania, the University of Portsmouth, 
SLAC National Accelerator Laboratory, Stanford University, the University of Sussex, and Texas A\&M University.

The DES data management system is supported by the National Science Foundation under Grant Number AST-1138766.
The DES participants from Spanish institutions are partially supported by MINECO under grants AYA2012-39559, ESP2013-48274, FPA2013-47986, and Centro de Excelencia Severo Ochoa SEV-2012-0234.
Research leading to these results has received funding from the European Research Council under the European Union’s Seventh Framework Programme (FP7/2007-2013) including ERC grant agreements 
 240672, 291329, and 306478.

Based in part on observations taken at the Australian Astronomical Observatory under program A/2013B/012.

Funding for SDSS-III has been provided by the Alfred P. Sloan Foundation, the
Participating Institutions, the National Science Foundation, and the
U.S. Department of Energy Office of Science. The SDSS-III web site is
http://www.sdss3.org/.

SDSS-III is managed by the Astrophysical Research Consortium for the
Participating Institutions of the SDSS-III Collaboration including the
University of Arizona, the Brazilian Participation Group, Brookhaven National
Laboratory, University of Cambridge, Carnegie Mellon University, University of
Florida, the French Participation Group, the German Participation Group,
Harvard University, the Instituto de Astrofisica de Canarias, the Michigan
State/Notre Dame/JINA Participation Group, Johns Hopkins University, Lawrence
Berkeley National Laboratory, Max Planck Institute for Astrophysics, Max Planck
Institute for Extraterrestrial Physics, New Mexico State University, New York
University, Ohio State University, Pennsylvania State University, University of
Portsmouth, Princeton University, the Spanish Participation Group, University
of Tokyo, University of Utah, Vanderbilt University, University of Virginia,
University of Washington, and Yale University.

This publication makes use of data products from the Two Micron All Sky Survey,
which is a joint project of the University of Massachusetts and the Infrared
Processing and Analysis Center/California Institute of Technology, funded by
the National Aeronautics and Space Administration and the National Science
Foundation.

GAMA is a joint European-Australasian project based around a spectroscopic
campaign using the Anglo-Australian Telescope. The GAMA input catalogue is
based on data taken from the Sloan Digital Sky Survey and the UKIRT Infrared
Deep Sky Survey. Complementary imaging of the GAMA regions is being obtained by
a number of independent survey programmes including GALEX MIS, VST KiDS, VISTA
VIKING, WISE, Herschel-ATLAS, GMRT and ASKAP providing UV to radio
coverage. GAMA is funded by the STFC (UK), the ARC (Australia), the AAO, and
the participating institutions. The GAMA website is http://www.gama-survey.org/
.

\appendix

\section{X-Ray Clusters}

The \emph{Chandra} clusters from Section~\ref{sec:chandra} are described in
Table~\ref{tab:chandraclusters}, and the XCS clusters from
Section~\ref{sec:xcs} are described in Table~\ref{tab:xcsclusters}.

\begin{deluxetable*}{llllllllll}
\tablewidth{0pt}
\tablecaption{\emph{Chandra} Clusters}
\tablehead{
  \colhead{Name} &
  \colhead{ID\tablenotemark{a}} &
  \colhead{$\lambda$} &
  \colhead{$\zlambda$} &
  \colhead{$\alpha_{\mathrm{CG}}$} &
  \colhead{$\delta_{\mathrm{CG}}$} &
  \colhead{$\alpha_{X}$} &
  \colhead{$\delta_{X}$} &
  \colhead{$T_X$} &
  \colhead{Notes}
}
\startdata
CXOU J224845.1-443144 &     2 & $174.7 \pm  5.2$ & $ 0.372 \pm  0.009$ &   342.237888 &   -44.502977 &   342.187790 &   -44.528860 &  $15.35^{+0.73}_{-0.45}$ & \\
CXOU J051635.7-543042 &     4 & $192.1 \pm  5.9$ & $ 0.325 \pm  0.016$ &    79.154972 &   -54.516379 &    79.148935 &   -54.511790 &  $11.24^{+0.83}_{-0.84}$ & 1,2\\
CXOU J004050.0-440757 &     8 & $143.0 \pm  7.7$ & $ 0.366 \pm  0.009$ &    10.208206 &   -44.130624 &    10.208210 &   -44.132540 &  $ 7.83^{+1.03}_{-0.77}$ & 1\\
CXOU J042605.0-545505 &    20 & $ 91.9 \pm  4.5$ & $ 0.642 \pm  0.011$ &    66.517163 &   -54.925298 &    66.520710 &   -54.918000 &  $ 7.57^{+1.98}_{-1.53}$ & \\
CXOU J045628.4-511640 &    38 & $ 91.6 \pm  5.1$ & $ 0.569 \pm  0.007$ &    74.117138 &   -51.276405 &    74.118490 &   -51.277660 &  $ 9.80^{+1.80}_{-1.05}$ & \\
CXOU J044148.1-485521 &    45 & $ 89.3 \pm  4.8$ & $ 0.812 \pm  0.012$ &    70.449577 &   -48.923361 &    70.450580 &   -48.922623 &  $ 7.19^{+1.06}_{-0.80}$ & 1\\
CXOU J044905.8-490131 &    54 & $ 92.8 \pm  4.9$ & $ 0.800 \pm  0.012$ &    72.266860 &   -49.027566 &    72.274050 &   -49.025320 &  $ 7.33^{+1.61}_{-0.97}$ & 1\\
CXOU J041804.1-475001 &   143 & $ 52.6 \pm  3.4$ & $ 0.584 \pm  0.007$ &    64.523720 &   -47.827636 &    64.516980 &   -47.833660 &  -- & \\
CXOU J095736.6+023427 &   183 & $ 58.4 \pm  4.6$ & $ 0.381 \pm  0.009$ &   149.404209 &     2.573747 &   149.402610 &     2.574050 &  $ 6.61^{+0.72}_{-0.64}$ & 2\\
CXOU J045314.4-594426 &   211 & $ 46.8 \pm  4.4$ & $ 0.315 \pm  0.018$ &    73.336516 &   -59.723625 &    73.310150 &   -59.740450 &  -- & \\
CXOU J043939.5-542420 &   260 & $ 58.7 \pm  4.1$ & $ 0.682 \pm  0.015$ &    69.916567 &   -54.403846 &    69.914690 &   -54.405470 &  $ 4.89^{+4.61}_{-1.47}$ & \\
CXOU J050921.2-534211 &   269 & $ 55.2 \pm  4.2$ & $ 0.461 \pm  0.009$ &    77.371828 &   -53.707888 &    77.338340 &   -53.703120 &  $ 9.54^{+1.52}_{-0.92}$ & \\
CXOU J044646.2-483336 &   353 & $ 48.5 \pm  3.9$ & $ 0.773 \pm  0.014$ &    71.693121 &   -48.558086 &    71.692480 &   -48.560120 &  -- & \\
CXOU J095902.5+025534 &   380 & $ 42.7 \pm  4.0$ & $ 0.366 \pm  0.011$ &   149.761335 &     2.929103 &   149.760330 &     2.926170 &  $ 4.03^{+0.65}_{-0.59}$ & 2\\
CXOU J100047.6+013940 &   388 & $ 29.8 \pm  2.4$ & $ 0.209 \pm  0.005$ &   150.189817 &     1.657398 &   150.198335 &     1.661128 &  $ 3.49^{+0.17}_{-0.16}$ & 2\\
CXOU J042741.7-544559 &   516 & $ 49.1 \pm  4.1$ & $ 0.435 \pm  0.010$ &    66.900538 &   -54.768035 &    66.923670 &   -54.766510 &  $ 2.84^{+1.46}_{-0.70}$ & \\
CXOU J045232.9-594528 &   578 & $ 28.1 \pm  2.8$ & $ 0.266 \pm  0.015$ &    73.072468 &   -59.741317 &    73.137020 &   -59.757810 &  -- & \\
CXOU J065638.9-555819 &   767 & $ 28.9 \pm  3.0$ & $ 0.269 \pm  0.015$ &   104.145859 &   -55.977785 &   104.161980 &   -55.972045 &  -- & \\
CXOU J034031.0-284834 &  1054 & $ 23.2 \pm  2.6$ & $ 0.475 \pm  0.011$ &    55.129143 &   -28.817229 &    55.129200 &   -28.809310 &  -- & \\
CXOU J010258.1-493019 &  1156 & $ 26.4 \pm  2.5$ & $ 0.711 \pm  0.022$ &    15.701349 &   -49.511298 &    15.741875 &   -49.505220 &  -- & \\
CXOU J004137.9-440225 &  1227 & $ 26.7 \pm  2.9$ & $ 0.459 \pm  0.011$ &    10.403128 &   -44.040263 &    10.407830 &   -44.040270 &  -- & \\
CXOU J003309.7-434745 &  1245 & $ 22.6 \pm  2.5$ & $ 0.407 \pm  0.010$ &     8.282083 &   -43.799248 &     8.290260 &   -43.795940 &  -- & 2\\
CXOU J045628.1-454024 &  1371 & $ 20.3 \pm  2.1$ & $ 0.578 \pm  0.009$ &    74.110967 &   -45.672684 &    74.117070 &   -45.673420 &  -- & \\
CXOU J022428.2-041529 &  1474 & $ 21.1 \pm  2.3$ & $ 0.254 \pm  0.013$ &    36.138450 &    -4.238674 &    36.117590 &    -4.258110 &  -- & 2\\
CXOU J034107.3-284559 &  1527 & $ 21.4 \pm  2.1$ & $ 0.589 \pm  0.011$ &    55.286213 &   -28.774285 &    55.280510 &   -28.766290 &  -- & \\
CXOU J100107.1+013408 &  1635 & $ 29.0 \pm  3.4$ & $ 0.381 \pm  0.013$ &   150.298618 &     1.554297 &   150.279780 &     1.569010 &  -- & 2\\
CXOU J022018.9-055647 &  1775 & $ 26.7 \pm  3.1$ & $ 0.660 \pm  0.018$ &    35.085095 &    -5.950116 &    35.078740 &    -5.946320 &  -- & \\
CXOU J044245.8-485443 &  1919 & $ 24.4 \pm  2.8$ & $ 0.820 \pm  0.015$ &    70.692931 &   -48.912217 &    70.690843 &   -48.911910 &  -- & \\
CXOU J044833.6-485007 &  1976 & $ 20.8 \pm  2.2$ & $ 0.421 \pm  0.021$ &    72.138634 &   -48.836412 &    72.140030 &   -48.835250 &  $ 5.59^{+4.24}_{-1.98}$ & \\
CXOU J044736.8-584530 &  2114 & $ 24.0 \pm  2.6$ & $ 0.681 \pm  0.020$ &    71.878993 &   -58.756044 &    71.903350 &   -58.758450 &  -- & \\
CXOU J095835.9+021235 &  2312 & $ 25.3 \pm  3.5$ & $ 0.944 \pm  0.017$ &   149.649663 &     2.209287 &   149.649662 &     2.209640 &  -- & \\
CXOU J045240.1-531552 &  2387 & $ 23.1 \pm  2.8$ & $ 0.687 \pm  0.024$ &    73.169362 &   -53.263914 &    73.167080 &   -53.264510 &  -- & \\
CXOU J095957.5+021825 &  2453 & $ 23.5 \pm  2.9$ & $ 0.923 \pm  0.016$ &   149.987795 &     2.315731 &   149.989540 &     2.306938 &  -- & \\
CXOU J100158.5+020352 &  2883 & $ 23.5 \pm  3.4$ & $ 0.441 \pm  0.012$ &   150.490085 &     2.069402 &   150.493733 &     2.064392 &  -- & \\
CXOU J045553.2-510748 &  3145 & $ 24.5 \pm  3.0$ & $ 0.756 \pm  0.024$ &    73.971873 &   -51.129557 &    73.971810 &   -51.130120 &  -- & \\

\enddata
\tablenotetext{a}{ID in expanded SVA1 catalog.}
\tablenotetext{1}{Also in SPT catalog~(S15).}
\tablenotetext{2}{Also in XCS catalog~(see Table~\ref{tab:xcsclusters}.}
\label{tab:chandraclusters}
\end{deluxetable*}

\begin{deluxetable*}{llllllllll}
\tablewidth{0pt}
\tablecaption{XCS Clusters}
\tablehead{
  \colhead{Name} &
  \colhead{ID\tablenotemark{a}} &
  \colhead{$\lambda$} &
  \colhead{$\zlambda$} &
  \colhead{$\alpha_{\mathrm{CG}}$} &
  \colhead{$\delta_{\mathrm{CG}}$} &
  \colhead{$\alpha_{X}$} &
  \colhead{$\delta_{X}$} &
  \colhead{$T_X$} &
  \colhead{Notes}
}
\startdata
XMMXCS J065828.8-555640.8 &     1 & $281.2 \pm  6.5$ & $ 0.298 \pm  0.017$ &   104.646822 &   -55.949043 &   104.620400 &   -55.944680 &  $ 9.44^{+0.14}_{-0.14}$ & \\
XMMXCS J051636.6-543120.8 &     4 & $192.1 \pm  5.9$ & $ 0.325 \pm  0.016$ &    79.154972 &   -54.516379 &    79.152740 &   -54.522467 &  $ 6.08^{+0.10}_{-0.10}$ & 1,2\\
XMMXCS J021441.2-043313.8 &    29 & $ 56.3 \pm  3.6$ & $ 0.139 \pm  0.004$ &    33.671242 &    -4.567278 &    33.671952 &    -4.553851 &  -- & \\
XMMXCS J044956.6-444017.3 &    32 & $ 55.4 \pm  2.6$ & $ 0.144 \pm  0.003$ &    72.485352 &   -44.673356 &    72.486117 &   -44.671479 &  -- & 1\\
XMMXCS J233227.2-535828.2 &    33 & $ 82.8 \pm  4.1$ & $ 0.424 \pm  0.007$ &   353.114476 &   -53.974433 &   353.113450 &   -53.974510 &  -- & \\
XMMXCS J095940.7+023110.8 &    74 & $ 81.5 \pm  4.9$ & $ 0.707 \pm  0.015$ &   149.923436 &     2.525051 &   149.919750 &     2.519675 &  $ 5.01^{+0.66}_{-0.54}$ & \\
XMMXCS J034005.2-285024.4 &    99 & $ 69.6 \pm  4.8$ & $ 0.344 \pm  0.014$ &    55.029953 &   -28.844377 &    55.021691 &   -28.840115 &  -- & \\
XMMXCS J224824.7-444225.3 &   136 & $ 63.0 \pm  4.3$ & $ 0.476 \pm  0.010$ &   342.098778 &   -44.708732 &   342.103250 &   -44.707049 &  -- & \\
XMMXCS J232956.5-560802.7 &   164 & $ 50.1 \pm  3.1$ & $ 0.418 \pm  0.010$ &   352.472225 &   -56.136006 &   352.485700 &   -56.134094 &  -- & \\
XMMXCS J095737.1+023428.9 &   183 & $ 58.4 \pm  4.6$ & $ 0.381 \pm  0.009$ &   149.404209 &     2.573747 &   149.404960 &     2.574713 &  $4.61^{0.59}_{-0.48}$ & 2\\
XMMXCS J003428.0-431854.2 &   274 & $ 49.5 \pm  5.1$ & $ 0.393 \pm  0.010$ &     8.614189 &   -43.316563 &     8.617005 &   -43.315066 &  $ 3.14^{+0.15}_{-0.14}$ & \\
XMMXCS J021734.7-051327.6 &   277 & $ 46.3 \pm  3.3$ & $ 0.658 \pm  0.014$ &    34.394127 &    -5.220327 &    34.394879 &    -5.224348 &  -- & \\
XMMXCS J045506.0-532342.4 &   299 & $ 41.8 \pm  2.8$ & $ 0.418 \pm  0.010$ &    73.773464 &   -53.396441 &    73.775354 &   -53.395126 &  $ 2.95^{+0.51}_{-0.40}$ & \\
XMMXCS J022511.8-062300.7 &   306 & $ 31.4 \pm  2.2$ & $ 0.215 \pm  0.005$ &    36.301178 &    -6.383116 &    36.299332 &    -6.383549 &  $ 2.38^{+0.55}_{-0.37}$ & \\
XMMXCS J095902.7+025544.9 &   380 & $ 42.7 \pm  4.0$ & $ 0.366 \pm  0.011$ &   149.761335 &     2.929103 &   149.761390 &     2.929155 &  $ 2.16^{+0.35}_{-0.29}$ & 2\\
XMMXCS J100047.3+013927.8 &   388 & $ 29.8 \pm  2.4$ & $ 0.209 \pm  0.005$ &   150.189817 &     1.657398 &   150.197330 &     1.657734 &  $ 3.18^{+0.16}_{-0.15}$ & 2\\
XMMXCS J233345.8-553826.9 &   451 & $ 45.6 \pm  3.6$ & $ 0.746 \pm  0.019$ &   353.441511 &   -55.637993 &   353.441180 &   -55.640811 &  $ 2.70^{+1.51}_{-0.89}$ & \\
XMMXCS J003346.3-431729.7 &   489 & $ 29.1 \pm  2.9$ & $ 0.214 \pm  0.005$ &     8.443268 &   -43.291959 &     8.442920 &   -43.291608 &  $ 2.49^{+0.13}_{-0.12}$ & \\
XMMXCS J232810.2-555015.8 &   889 & $ 40.1 \pm  3.6$ & $ 0.813 \pm  0.014$ &   352.031286 &   -55.839880 &   352.042820 &   -55.837728 &  -- & \\
XMMXCS J095901.2+024740.4 &  1193 & $ 20.4 \pm  2.4$ & $ 0.504 \pm  0.012$ &   149.756320 &     2.794723 &   149.755310 &     2.794571 &  -- & \\
XMMXCS J233000.5-543706.3 &  1198 & $ 20.6 \pm  1.9$ & $ 0.176 \pm  0.004$ &   352.501689 &   -54.618800 &   352.502360 &   -54.618431 &  $ 2.27^{+0.17}_{-0.15}$ & \\
XMMXCS J003309.8-434758.3 &  1245 & $ 22.6 \pm  2.5$ & $ 0.407 \pm  0.010$ &     8.282083 &   -43.799248 &     8.290958 &   -43.799532 &  -- & 2\\
XMMXCS J022827.3-042538.7 &  1434 & $ 23.6 \pm  2.5$ & $ 0.434 \pm  0.014$ &    37.115911 &    -4.435404 &    37.114008 &    -4.427436 &  $ 3.88^{+1.05}_{-0.71}$ & \\
XMMXCS J022433.9-041432.7 &  1474 & $ 21.1 \pm  2.3$ & $ 0.254 \pm  0.013$ &    36.138450 &    -4.238674 &    36.141298 &    -4.242430 &  $ 1.36^{+0.10}_{-0.08}$ & 2\\
XMMXCS J100109.1+013336.8 &  1635 & $ 29.0 \pm  3.4$ & $ 0.381 \pm  0.013$ &   150.298618 &     1.554297 &   150.288320 &     1.560238 &  -- & 2\\
XMMXCS J022307.9-041257.2 &  1707 & $ 20.0 \pm  2.0$ & $ 0.618 \pm  0.013$ &    35.794975 &    -4.214364 &    35.782951 &    -4.215907 &  -- & \\
XMMXCS J003627.6-432830.3 &  1868 & $ 23.5 \pm  2.7$ & $ 0.397 \pm  0.015$ &     9.109958 &   -43.453131 &     9.115160 &   -43.475104 &  -- & \\
XMMXCS J021755.3-052708.0 &  2833 & $ 21.8 \pm  2.7$ & $ 0.667 \pm  0.021$ &    34.475702 &    -5.451563 &    34.480539 &    -5.452240 &  -- & \\
XMMXCS J033931.8-283444.7 &  5590 & $ 21.1 \pm  3.1$ & $ 0.824 \pm  0.015$ &    54.901800 &   -28.575329 &    54.882578 &   -28.579090 &  -- & \\

\enddata
\tablenotetext{a}{ID in expanded SVA1 catalog.}
\tablenotetext{1}{Also in SPT catalog.}
\tablenotetext{2}{Also in \emph{Chandra} catalog.}
\label{tab:xcsclusters}
\end{deluxetable*}

\section{Data Catalog Formats}
\label{app:catalogs}

The full \redmapper{} SDSS DR8 and DES SVA1 catalogs will be available at {\tt
  http://risa.stanford.edu/redmapper/} in FITS format, and the DES SVA1
catalogs at {\tt http://des.ncsa.illinois.edu/releases/sva1}.  The catalogs
will also be available from the online journal in machine-readable formats.  A
summary of all the data tables provided is shown in Table~\ref{tab:allcats},
with pointers to the associated tables which describe the data products.  Note
that there are two versions of the SVA1 catalog -- the fiducial catalog, and
the expanded-footprint catalog with inferior star/galaxy separation and less
aggressive masking.  The cluster ID numbers are not matched between these two
versions of the catalog, which are considered distinct.

\begin{deluxetable*}{lll}
\tablewidth{0pt}
\tablecaption{\redmapper{} Catalogs and Associated Products}
\tablehead{
  \colhead{Filename} &
  \colhead{Description} &
  \colhead{Table Reference}
}
\startdata
{\tt redmapper\_dr8\_public\_v6.3\_catalog.fits} & SDSS DR8 catalog &
Table~\ref{tab:dr8cat}\\
{\tt redmapper\_dr8\_public\_v6.3\_members.fits} & SDSS DR8 members &
Table~\ref{tab:dr8mem}\\
{\tt redmapper\_dr8\_public\_v6.3\_zmask.fits} & SDSS DR8 $\zmax$ map & Table~\ref{tab:zmaskcat}\\
{\tt redmapper\_dr8\_public\_v6.3\_randoms.fits} & SDSS DR8 random points &
Table~\ref{tab:randcat}\\
{\tt redmapper\_dr8\_public\_v6.3\_area.fits} & SDSS DR8 effective area & Table~\ref{tab:effarea}\\
{\tt redmapper\_sva1\_public\_v6.3\_catalog.fits} & DES SVA1 catalog &
Table~\ref{tab:sva1cat}\\
{\tt redmapper\_sva1\_public\_v6.3\_members.fits} & DES SVA1 members &
Table~\ref{tab:sva1mem}\\
{\tt redmapper\_sva1-expanded\_public\_v6.3\_catalog.fits} & DES SVA1 expanded
catalog\tablenotemark{a} & Table~\ref{tab:sva1cat}\\
{\tt redmapper\_sva1-expanded\_public\_v6.3\_members.fits} & DES SVA1 expanded
members\tablenotemark{a} & Table~\ref{tab:sva1mem}\\
{\tt redmapper\_sva1\_public\_v6.3\_zmask.fits} & DES SVA1 $\zmax$ map & Table~\ref{tab:zmaskcat}\\
{\tt redmapper\_sva1\_public\_v6.3\_randoms.fits} & DES SVA1 random points &
Table~\ref{tab:randcat}\\
{\tt redmapper\_sva1\_public\_v6.3\_area.fits} & DES SVA1 effective area & Table~\ref{tab:effarea}\\
\tablenotetext{a}{See Section~\ref{sec:masking}.}
\enddata
\label{tab:allcats}

\end{deluxetable*}

\begin{deluxetable*}{lll}
  \tablewidth{0pt}
\tablecaption{\redmapper{} DR8 Cluster Catalog Format}
\tablehead{
  \colhead{Name} &
  \colhead{Data Type} &
  \colhead{Description}
}
\startdata
ID & INT(4) & \redmapper{} Cluster Identification Number\\
NAME & CHAR(20) & \redmapper{} Cluster Name\\
RA & FLOAT(8) & Right ascension in decimal degrees (J2000)\\
DEC & FLOAT(8) & Declination in decimal degrees (J2000)\\
Z\_LAMBDA & FLOAT(4) & Cluster \photoz $\zlambda$\\
Z\_LAMBDA\_ERR & FLOAT(4) & Gaussian error estimate for $\zlambda$\\
LAMBDA & FLOAT(4) & Richness estimate $\lambda$\\
LAMBDA\_ERR & FLOAT(4) & Gaussian error estimate for $\lambda$\\
S & FLOAT(4) & Richness scale factor (see Eqn.~\ref{eqn:S})\\
Z\_SPEC & FLOAT(4) & SDSS spectroscopic redshift for most likely center (-1.0 if not available)\\
OBJID & INT(8) & SDSS DR8 CAS object identifier\\
IMAG & FLOAT(4) & $i$-band cmodel magnitude for most likely central galaxy (dereddened)\\
IMAG\_ERR & FLOAT(4) & error on $i$-band cmodel magnitude\\
MODEL\_MAG\_U & FLOAT(4) & $u$ model magnitude for most likely central galaxy
(dereddened)\\
MODEL\_MAGERR\_U & FLOAT(4) & error on $u$ model magnitude\\
MODEL\_MAG\_G & FLOAT(4) & $g$ model magnitude for most likely central galaxy
(dereddened)\\
MODEL\_MAGERR\_G & FLOAT(4) & error on $g$ model magnitude\\
MODEL\_MAG\_R & FLOAT(4) & $r$ model magnitude for most likely central galaxy
(dereddened)\\
MODEL\_MAGERR\_R & FLOAT(4) & error on $r$ model magnitude\\
MODEL\_MAG\_I & FLOAT(4) & $i$ model magnitude for most likely central galaxy (dereddened)\\
MODEL\_MAGERR\_I & FLOAT(4) & error on $i$ model magnitude\\
MODEL\_MAG\_Z & FLOAT(4) & $z$ model magnitude for most likely central galaxy
(dereddened)\\
MODEL\_MAGERR\_Z & FLOAT(4) & error on $z$ model magnitude\\
ILUM & FLOAT(4) & Total membership-weighted $i$-band luminosity (units of
$L_*$)\\
P\_CEN[5] & 5$\times$FLOAT(4) & Centering probability $\Pcen$ for 5 most likely centrals\\
RA\_CEN[5] & 5$\times$FLOAT(8) & R.A. for 5 most likely centrals\\
DEC\_CEN[5] & 5$\times$FLOAT(8) & Decl. for 5 most likely centrals\\
ID\_CEN[5] & 5$\times$INT(8) & DR8 CAS object identifier for 5 most likely centrals\\
PZBINS[21] & 21$\times$FLOAT(4) & Redshift points at which $P(z)$ is evaluated\\
PZ[21] & 21$\times$FLOAT(4) & $P(z)$ evaluated at redshift points given by PZBINS\\
\enddata
\label{tab:dr8cat}
\end{deluxetable*}

\begin{deluxetable*}{lll}
\tablewidth{0pt}
\tablecaption{\redmapper{} DR8 Member Catalog Format}
\tablehead{
  \colhead{Name} &
  \colhead{Format} &
  \colhead{Description}
}
\startdata
ID & INT(4) & \redmapper{} Cluster Identification Number\\
RA & FLOAT(8) & Right ascension in decimal degrees (J2000)\\
DEC & FLOAT(8) & Declination in decimal degrees (J2000)\\
R & FLOAT(4) & Distance from cluster center ($h^{-1}\,\mathrm{Mpc}$)\\
P & FLOAT(4) & Membership probability\\
P\_FREE & FLOAT(4) & Probability that member is not a member of a higher ranked
cluster\\
THETA\_L & FLOAT(4) & Luminosity ($i$-band) weight\\
THETA\_R & FLOAT(4) & Radial weight\\
IMAG & FLOAT(4) & $i$-band cmodel magnitude (dereddened)\\
IMAG\_ERR & FLOAT(4) & error on $i$-band cmodel magnitude\\
MODEL\_MAG\_U & FLOAT(4) & $u$ model magnitude (dereddened)\\
MODEL\_MAGERR\_U & FLOAT(4) & error on $u$ model magnitude\\
MODEL\_MAG\_G & FLOAT(4) & $g$ model magnitude (dereddened)\\
MODEL\_MAGERR\_G & FLOAT(4) & error on $g$ model magnitude\\
MODEL\_MAG\_R & FLOAT(4) & $r$ model magnitude (dereddened)\\
MODEL\_MAGERR\_R & FLOAT(4) & error on $r$ model magnitude\\
MODEL\_MAG\_I & FLOAT(4) & $i$ model magnitude (dereddened)\\
MODEL\_MAGERR\_I & FLOAT(4) & error on $i$ model magnitude\\
MODEL\_MAG\_Z & FLOAT(4) & $z$ model magnitude (dereddened)\\
MODEL\_MAGERR\_Z & FLOAT(4) & error on $z$ model magnitude\\
Z\_SPEC & FLOAT(4) & SDSS spectroscopic redshift (-1.0 if not available)\\
OBJID & INT(8) &  SDSS DR8 CAS object identifier\\
\enddata
\label{tab:dr8mem}
\tablecomments{The probability $p$ is the raw membership probability, while the
  probability $\pfree$ is the probability that the galaxy does not belong to a
  previous cluster in the percolation.  The total membership probability must
  be modified by the radial and luminosity weights, such that $\pmem = p \times
  \pfree \times \theta_i \times \theta_r$.}
\end{deluxetable*}

\begin{deluxetable*}{lll}
  \tablewidth{0pt}
\tablecaption{\redmapper{} SVA1 Cluster Catalog Format}
\tablehead{
  \colhead{Name} &
  \colhead{Format} &
  \colhead{Description}
}
\startdata
ID & INT(4) & \redmapper{} Cluster Identification Number\\
NAME & CHAR(20) & \redmapper{} Cluster Name\\
RA & FLOAT(8) & Right ascension in decimal degrees (J2000)\\
DEC & FLOAT(8) & Declination in decimal degrees (J2000)\\
Z\_LAMBDA & FLOAT(4) & Cluster \photoz $\zlambda$\\
Z\_LAMBDA\_ERR & FLOAT(4) & Gaussian error estimate for $\zlambda$\\
LAMBDA & FLOAT(4) & Richness estimate $\lambda$\\
LAMBDA\_ERR & FLOAT(4) & Gaussian error estimate for $\lambda$\\
S & FLOAT(4) & Richness scale factor (see Eqn.~\ref{eqn:S})\\
Z\_SPEC & FLOAT(4) & SDSS spectroscopic redshift for most likely center (-1.0 if not available)\\
COADD\_OBJECTS\_ID & INT(8) & DES {\tt COADD\_OBJECTS\_ID} identification number \\
MAG\_AUTO\_G & FLOAT(4) & $g$ MAG\_AUTO magnitude for most likely central
galaxy (SLR corrected)\\
MAGERR\_AUTO\_G & FLOAT(4) & error on $g$ MAG\_AUTO magnitude\\
MAG\_AUTO\_R & FLOAT(4) & $r$ MAG\_AUTO magnitude for most likely central
galaxy (SLR corrected)\\
MAGERR\_AUTO\_R & FLOAT(4) & error on $g$ MAG\_AUTO magnitude\\
MAG\_AUTO\_I & FLOAT(4) & $i$ MAG\_AUTO magnitude for most likely central
galaxy (SLR corrected)\\
MAGERR\_AUTO\_I & FLOAT(4) & error on $g$ MAG\_AUTO magnitude\\
MAG\_AUTO\_Z & FLOAT(4) & $z$ MAG\_AUTO magnitude for most likely central
galaxy (SLR corrected)\\
MAGERR\_AUTO\_Z & FLOAT(4) & error on $g$ MAG\_AUTO magnitude\\
ZLUM & FLOAT(4) & Total membership-weighted $z$-band luminosity (units of
$L_*$)\\
P\_CEN[5] & 5$\times$FLOAT(4) & Centering probability $\Pcen$ for 5 most likely centrals\\
RA\_CEN[5] & 5$\times$FLOAT(8) & R.A. for 5 most likely centrals\\
DEC\_CEN[5] & 5$\times$FLOAT(8) & Decl. for 5 most likely centrals\\
ID\_CEN[5] & 5$\times$INT(8) & DES {\tt COADD\_OBJECTS\_ID} identification number for
5 most likely centrals\\
PZBINS[21] & 21$\times$FLOAT(4) & Redshift points at which $P(z)$ is evaluated\\
PZ[21] & 21$\times$FLOAT(4) & $P(z)$ evaluated at redshift points given by PZBINS\\
\enddata
\label{tab:sva1cat}
\end{deluxetable*}

\begin{deluxetable*}{lll}
\tablewidth{0pt}
\tablecaption{\redmapper{} DES SVA1 Member Catalog Format}
\tablehead{
  \colhead{Name} &
  \colhead{Format} &
  \colhead{Description}
}
\startdata
ID & INT(4) & \redmapper{} Cluster Identification Number\\
RA & FLOAT(8) & Right ascension in decimal degrees (J2000)\\
DEC & FLOAT(8) & Declination in decimal degrees (J2000)\\
R & FLOAT(4) & Distance from cluster center ($h^{-1}\,\mathrm{Mpc}$)\\
P & FLOAT(4) & Membership probability\\
P\_FREE & FLOAT(4) & Probability that member is not a member of a higher ranked
cluster\\
THETA\_L & FLOAT(4) & Luminosity ($z$-band) weight\\
THETA\_R & FLOAT(4) & Radial weight\\
MAG\_AUTO\_G & FLOAT(4) & $g$ MAG\_AUTO magnitude (SLR corrected)\\
MAGERR\_AUTO\_G & FLOAT(4) & error on $g$ MAG\_AUTO magnitude\\
MAG\_AUTO\_R & FLOAT(4) & $r$ MAG\_AUTO magnitude (SLR corrected)\\
MAGERR\_AUTO\_R & FLOAT(4) & error on $r$ MAG\_AUTO magnitude\\
MAG\_AUTO\_I & FLOAT(4) & $i$ MAG\_AUTO magnitude (SLR corrected)\\
MAGERR\_AUTO\_I & FLOAT(4) & error on $i$ MAG\_AUTO magnitude\\
MAG\_AUTO\_Z & FLOAT(4) & $z$ MAG\_AUTO magnitude (SLR corrected)\\
MAGERR\_AUTO\_Z & FLOAT(4) & error on $z$ MAG\_AUTO magnitude\\
Z\_SPEC & FLOAT(4) & Spectroscopic redshift (-1.0 if not available)\\
COADD\_OBJECTS\_ID & INT(8) & DES {\tt COADD\_OBJECTS\_ID} identification number\\
\enddata
\label{tab:sva1mem}
\tablecomments{See Table~\ref{tab:dr8mem} for information on how to compute $\pmem$.}
\end{deluxetable*}

\begin{deluxetable*}{lll}
\tablewidth{0pt}
\tablecaption{\redmapper{} $\zmax$ Map Format}
\tablehead{
  \colhead{Name} &
  \colhead{Format} &
  \colhead{Description}
}
\startdata
HPIX\tablenotemark{a} & INT(8) & \healpix{} ring-ordered pixel number\\
ZMAX & FLOAT(4) & Maximum redshift of a cluster centered in this pixel\\
FRACGOOD & FLOAT(4) & Fraction of pixel area that is not masked\\
\enddata
\tablenotetext{a}{We use {\tt NSIDE=4096} for the SVA1 catalogs, and {\tt
    NSIDE=2048} for the DR8 catalog.}
\label{tab:zmaskcat}
\end{deluxetable*}

\begin{deluxetable*}{lll}
\tablewidth{0pt}
\tablecaption{\redmapper{} Random Points Catalog Format}
\tablehead{
  \colhead{Name} &
  \colhead{Format} &
  \colhead{Description}
}
\startdata
RA & FLOAT(8) & Right ascension in decimal degrees (J2000)\\
DEC & FLOAT(8) & Declination in decimal degrees (J2000)\\
Z & FLOAT(4) & Redshift of random point\\
LAMBDA & FLOAT(4) & Richness of random point\\
WEIGHT & FLOAT(4) & Weight of random point\\
\enddata
\label{tab:randcat}
\end{deluxetable*}

\begin{deluxetable*}{lll}
\tablewidth{0pt}
\tablecaption{\redmapper{} Effective Area Format}
\tablehead{
  \colhead{Name} &
  \colhead{Format} &
  \colhead{Description}
}
\startdata
Z & FLOAT(4) & Redshift cut\\
AREA & FLOAT(4) & Effective area\\
\enddata
\label{tab:effarea}
\end{deluxetable*}

\section{Affiliations}
\small{
\noindent$^{1}$ Kavli Institute for Particle Astrophysics \& Cosmology, P. O. Box 2450, Stanford University, Stanford, CA 94305, USA\\
$^{2}$ SLAC National Accelerator Laboratory, Menlo Park, CA 94025, USA\\
$^{3}$ Department of Physics, University of Arizona, Tucson, AZ 85721, USA\\
$^{4}$ Department of Physics and Santa Cruz Institute for Particle Physics, University of California, Santa Cruz, CA 95064, USA\\
$^{5}$ Department of Physics and Astronomy, Pevensey Building, University of Sussex, Brighton, BN1 9QH, UK\\
$^{6}$ Faculty of Physics, Ludwig-Maximilians University, Scheinerstr. 1, 81679 Munich, Germany\\
$^{7}$ Department of Physics, Stanford University, 382 Via Pueblo Mall, Stanford, CA 94305, USA\\
$^{8}$ Institute of Cosmology \& Gravitation, University of Portsmouth, Portsmouth, PO1 3FX, UK\\
$^{9}$ Cerro Tololo Inter-American Observatory, National Optical Astronomy Observatory, Casilla 603, La Serena, Chile\\
$^{10}$ Department of Physics \& Astronomy, University College London, Gower Street, London, WC1E 6BT, UK\\
$^{11}$ Department of Physics and Electronics, Rhodes University, PO Box 94, Grahamstown, 6140, South Africa\\
$^{12}$ Fermi National Accelerator Laboratory, P. O. Box 500, Batavia, IL 60510, USA\\
$^{13}$ CNRS, UMR 7095, Institut d'Astrophysique de Paris, F-75014, Paris, France\\
$^{14}$ Sorbonne Universit\'es, UPMC Univ Paris 06, UMR 7095, Institut d'Astrophysique de Paris, F-75014, Paris, France\\
$^{15}$ Department of Physics and Astronomy, University of Pennsylvania, Philadelphia, PA 19104, USA\\
$^{16}$ Laborat\'orio Interinstitucional de e-Astronomia - LIneA, Rua Gal. Jos\'e Cristino 77, Rio de Janeiro, RJ - 20921-400, Brazil\\
$^{17}$ Observat\'orio Nacional, Rua Gal. Jos\'e Cristino 77, Rio de Janeiro, RJ - 20921-400, Brazil\\
$^{18}$ Department of Astronomy, University of Illinois, 1002 W. Green Street, Urbana, IL 61801, USA\\
$^{19}$ National Center for Supercomputing Applications, 1205 West Clark St., Urbana, IL 61801, USA\\
$^{20}$ Institut de Ci\`encies de l'Espai, IEEC-CSIC, Campus UAB, Carrer de Can Magrans, s/n,  08193 Bellaterra, Barcelona, Spain\\
$^{21}$ ARC Centre of Excellence for All-sky Astrophysics (CAASTRO), Australian National University, Canberra, ACT 2611, Australia\\
$^{22}$ The Research School of Astronomy and Astrophysics, Australian National University, ACT 2601, Australia\\
$^{23}$ Astrophysics Research Institute, Liverpool John Moores University, IC2, Liverpool Science Park, Brownlow Hill, Liverpool, L5 3AF\\
$^{24}$ School of Physics and Astronomy, University of Southampton,  Southampton, SO17 1BJ, UK\\
$^{25}$ School of Mathematics and Physics, University of Queensland, Brisbane, QLD 4072, Australia\\
$^{26}$ Excellence Cluster Universe, Boltzmannstr.\ 2, 85748 Garching, Germany\\
$^{27}$ Department of Astronomy, University of Michigan, Ann Arbor, MI 48109, USA\\
$^{28}$ Department of Physics, University of Michigan, Ann Arbor, MI 48109, USA\\
$^{29}$ Kavli Institute for Cosmological Physics, University of Chicago, Chicago, IL 60637, USA\\
$^{30}$ Centre for Astrophysics \& Supercomputing, Swinburne University of Technology, Victoria 3122, Australia\\
$^{31}$ Department of Astronomy, University of California, Berkeley,  501 Campbell Hall, Berkeley, CA 94720, USA\\
$^{32}$ Lawrence Berkeley National Laboratory, 1 Cyclotron Road, Berkeley, CA 94720, USA\\
$^{33}$ Max Planck Institute for Extraterrestrial Physics, Giessenbachstrasse, 85748 Garching, Germany\\
$^{34}$ Universit\"ats-Sternwarte, Fakult\"at f\"ur Physik, Ludwig-Maximilians Universit\"at M\"unchen, Scheinerstr. 1, 81679 M\"unchen, Germany\\
$^{35}$ Astrophysics \& Cosmology Research Unit, School of Mathematics, Statistics \& Computer Science, University of KwaZulu-Natal, Westville Campus, Durban 4041, South Africa\\
$^{36}$ Center for Cosmology and Astro-Particle Physics, The Ohio State University, Columbus, OH 43210, USA\\
$^{37}$ Department of Physics, The Ohio State University, Columbus, OH 43210, USA\\
$^{38}$ Jodrell Bank Center for Astrophysics, School of Physics and Astronomy, University of Manchester, Oxford Road, Manchester, M13 9PL, UK\\
$^{39}$ Australian Astronomical Observatory, North Ryde, NSW 2113, Australia\\
$^{40}$ Sydney Institute for Astronomy, School of Physics, A28, The University of Sydney, NSW 2006, Australia\\
$^{41}$ Departamento de F\'{\i}sica Matem\'atica,  Instituto de F\'{\i}sica, Universidade de S\~ao Paulo,  CP 66318, CEP 05314-970, S\~ao Paulo, SP,  Brazil\\
$^{42}$ Institute for Astronomy, University of Edinburgh, Royal Observatory, Blackford Hill, Edinburgh, EH9 3HJ, UK\\
$^{43}$ George P. and Cynthia Woods Mitchell Institute for Fundamental Physics and Astronomy, and Department of Physics and Astronomy, Texas A\&M University, College Station, TX 77843,  USA\\
$^{44}$ Department of Astronomy, The Ohio State University, Columbus, OH 43210, USA\\
$^{45}$ Department of Astrophysical Sciences, Princeton University, Peyton Hall, Princeton, NJ 08544, USA\\
$^{46}$ Instituci\'o Catalana de Recerca i Estudis Avan\c{c}ats, E-08010 Barcelona, Spain\\
$^{47}$ Institut de F\'{\i}sica d'Altes Energies (IFAE), The Barcelona Institute of Science and Technology, Campus UAB, 08193 Bellaterra (Barcelona) Spain\\
$^{48}$ Jet Propulsion Laboratory, California Institute of Technology, 4800 Oak Grove Dr., Pasadena, CA 91109, USA\\
$^{49}$ BIPAC, Department of Physics, University of Oxford, Denys Wilkinson Building, 1 Keble Road, Oxford OX1 3RH, UK\\
$^{50}$ Centro de Investigaciones Energ\'eticas, Medioambientales y Tecnol\'ogicas (CIEMAT), Madrid, Spain\\
$^{51}$ Instituto de F\'\i sica, UFRGS, Caixa Postal 15051, Porto Alegre, RS - 91501-970, Brazil\\
$^{52}$ Instituto de Astrof\'{\i}sica e Ci\^{e}ncias do Espa\c{c}o, Universidade do Porto, CAUP, Rua das Estrelas, 4150-762 Porto, Portugal\\
$^{53}$ Departamento de F\'isica e Astronomia, Faculdade de Ci\^encias, Universidade do Porto, Rua do Campo Alegre, 687, 4169-007 Porto, Portugal\\
$^{54}$ Argonne National Laboratory, 9700 South Cass Avenue, Lemont, IL 60439, USA\\
}
\end{document}